\documentclass{aastex}          
\usepackage{spr-astr-addons}    

\usepackage{amsmath}
\usepackage{color}

\begin{document}

\title{Reassessing Dust's Role in Forming the CMB}

\shorttitle{Dust's Role in Forming the CMB}
\shortauthors{Melia}

\author{F. Melia\altaffilmark{1}}
\affil{Department of Physics, the Applied Math Program, and Department of Astronomy,
The University of Arizona, Tucson, AZ 85721 \\
E-mail: fmelia@email.arizona.edu}

\altaffiltext{1}{John Woodruff Simpson Fellow.} 

\begin{abstract}
The notion that dust might have formed the cosmic microwave background
(CMB) has been strongly refuted on the strength of four decades of
observation and analysis, in favour of recombination at a redshift
$z\sim 1080$. But tension with the data is growing in several other
areas, including measurements of the Hubble constant $H(z)$ and the
BAO scale, which directly or indirectly impact the physics at the
surface of last scattering (LSS). The $R_{\rm h}=ct$ universe
resolves at least some of this tension. We show in this paper that---{\it if
the BAO scale is in fact equal to the acoustic horizon}---the redshift
of the LSS in this cosmology is $z_{\rm cmb} \sim 16$, placing it within
the era of Pop III star formation, prior to the epoch of reionization
at $15\gtrsim z \gtrsim 6$. Quite remarkably, the measured values of
$z_{\rm cmb}$ and $H_0\equiv H(0)$ in this model are sufficient to
argue that the CMB temperature today ought to be $\sim 3$ K, so
$H_0$ and the baryon to photon ratio are not independent free
parameters. This scenario might have resulted from rethermalization
of the CMB photons by dust, presumably supplied to the interstellar medium
by the ejecta of Pop III stars. Dust rethermalization may therefore yet
resurface as a relevant ingredient in the $R_{\rm h}=ct$ universe. Upcoming
high sensitivity instruments should be able to readily distinguish between the
recombination and dust scenarios by either (i) detecting recombination lines
at $z\sim 1080$, or (ii) establishing a robust frequency-dependent variation
of the CMB power spectrum at the level of $\sim 2-4\%$ across the sampled
frequency range.
\end{abstract}


\section{Introduction}
Since COBE's discovery \citep{Mather1990} that the spectrum of the
cosmic microwave background (CMB) is a near perfect blackbody, all
succeeding measurements of this relic signal (including those reported in
refs.~\citealt{Hinshaw2003,Planck2014}) have been
interpreted self-consistently in terms of a model in which the radiation
was thermalized within one year of the big bang. Diffusing through a
gradually thinning, scattering-dominated medium, these photons eventually
streamed freely once the protons and electrons in the cosmic fluid combined
to form neutral hydrogen and helium, a process (not so accurately) referred
to as `recombination.'

The actual origin of the CMB was not always so evident, however, and
serious consideration had been given to the possibility that it was produced
by dust in the early Universe, injected into the interstellar medium (ISM) by
Pop III stars \citep{Rees1978,Rowan1979,Wright1982}.
An additional attraction of this scenario was the likelihood that the photons
rethermalized by dust were themselves emitted by the same stars, thereby closing
the loop on a potentially elegant, self-consistent physical picture.

But the dust model for the CMB very quickly gave ground to recombination
for several telling reasons. Two of them, in particular, relied heavily
on each other and suggested quite emphatically that the surface of last
scattering (LSS) had to lie at a redshift $z_{\rm cmb}\sim 1080$. First,
there was the inference of a characteristic scale in the CMB's power spectrum
\citep{Spergel2003} which, when identified as an acoustic horizon (see below),
implied that radiation must have decoupled from the baryonic fluid no more than
$\sim 380,000$ yrs after the big bang (placing it at the aforementioned
$z_{\rm cmb}\sim 1080$). Second, one could reasonably assume that the CMB propagated
more or less freely after this time, so that its temperature scaled as
$T(z)\propto (1+z)$. Assuming that the radiation and matter were in thermal
equilibrium prior to the LSS, one could then use the Saha equation to estimate
the temperature $T_{\rm cmb}$---and hence the redshift---at which the free electron
fraction dropped to $50\%$, signaling the time during which the baryonic fluid
transitioned from ionized plasma to neutral gas. Recombination would have
occurred at $T_{\rm cmb}\sim 3,000$ K and, given a measured CMB temperature today
of $\sim 2.728$ K, this would imply a redshift $z_{\rm cmb}\sim 1,100$, nicely consistent
with the interpretation of the acoustic scale. In contrast, emission dominated by
dust at $T_{\rm cmb}\lesssim 50$ K would have placed the redshift $z_{\rm cmb}$ at
no more than $\sim 20$, creating a significant conflict with the acoustic-scale
interpretation of the peaks in the CMB power spectrum. And in parallel
with such arguments for recombination, there was also growing concern that Pop III
starlight scattered by the stars' own ejected dust faced seemingly insurmountable
difficulties accounting for the observed CMB spectrum (see, e.g., Li 2003).

Today, there is very little doubt that the CMB must have formed
via recombination at $z\sim 1080$ in the context of $\Lambda$CDM. A dust
scenario would produce too many inconsistencies with the age-redshift
relation and the Pop III star formation rate, among many other observables.
As the precision and breadth of the measurements continued to improve,
however, the basic recombination picture for the CMB's origin has not
remained as clear as one might have hoped two decades ago---not because of
problems with the CMB itself but, rather, because of the tension this
interpretation creates with other kinds of cosmological observations.
For example, from the analysis of the CMB observed with {\it Planck}
(Planck Collaboration 2014), one infers a value of the Hubble constant
($H_0=67.6\pm0.9$ km s$^{-1}$ Mpc$^{-1}$) lower than is typically
measured locally, and a higher value for the matter fluctuation amplitude
($\sigma_8$) than is derived from Sunyaev-Zeldovich data. Quite tellingly,
none of the extensions to the six-parameter standard $\Lambda$CDM model
explored by the Planck team was able to resolve these inconsistencies.
As we shall see below, comparable tension now exists also between the baryon
acoustic oscillation (BAO) scale inferred from the galaxy and quasar distributions
at $z\sim 0.5-2.34$ and the aforementioned acoustic length seen in the CMB, weakening
the argument for an LSS at $z_{\rm cmb}\sim 1080$.

Over the past decade, the standard model's inability to resolve such tensions,
along with several inexplicable coincidences, have led to the development
of an alternative Friedmann-Robertson-Walker cosmology known as the $R_{\rm h}=ct$
universe \citep{Melia2007,Melia2016,Melia2017b,MeliaAbdelqader2009,MeliaShevchuk2012}.
During this time, the predictions of $R_{\rm h}=ct$ have been compared with
those of $\Lambda$CDM using over 23 different kinds of data, outperforming
the standard model in every case (see, e.g., Table I in Melia 2017a).
We are therefore motivated to consider how the origin of the CMB might be
interpreted in this alternative cosmology. Ironically, we shall find
that---if the BAO and acoustic scales are the same---the redshift of the
LSS in this model had to be $\sim 16$, remarkably close to what would have
been required in the original dust model. We shall also find that this
redshift sits right within the period of Pop III star formation, prior to
the epoch of reionization ($5\lesssim z \lesssim 15$), a likely
time during which dust would have been injected into the ISM. And
quite interestingly, we shall also determine that if this model is correct,
knowledge of $H_0$ and $z_{\rm cmb}$ by themselves is sufficient to argue that
the CMB temperature today should be $\sim 3$ K, very close to the actual value,
suggesting that the Hubble constant and the baryon to photon ratio are not independent,
free parameters.

Our goal in this paper is therefore not to critique the basic
recombination picture in $\Lambda$CDM which, as noted earlier, matches the data
remarkably well but, rather, to demonstrate how the (now dated) dust model for
the origin of the CMB may still be viable, albeit in the context of $R_{\rm h}=ct$.
The growing tension between the predictions of the standard model and the ever 
improving observations \citep{MeliaGenova2018} could certainly benefit from a 
reconsideration of a dust origin for the CMB. But our principal motivation for 
reanalyzing this mechanism is that, while recombination does not work for 
$R_{\rm h}=ct$, the dust model is unavoidable. It is our
primary goal to examine how and why this association  emerges naturally in this
cosmology. The analysis in this paper will show that, while dust reprocessing
of radiation emitted by the same first generation stars was part of the original
proposal, our improved understanding of star formation during the Pop III
era precludes this possibility. Instead, the
background radiation would have originated between the big bang and
decoupling, similarly to the situation in $\Lambda$CDM, but would have
been reprocessed by dust prior to reionization in the context of $R_{\rm h}=ct$.
A critical difference between these models is that the anisotropies
in the observed CMB field would therefore correspond to large-scale structure
at $z\sim 16$ in $R_{\rm h}=ct$, instead of $z\sim 1080$ in the standard
picture.

There are, of course, several definitive tests one may carry out to
distinguish between these two scenarios, and we shall consider them in our analysis,
described in detail in \S~VI. In this section, we shall also describe several potential
shortcomings of a dusty origin for the CMB versus the current recombination
picture, and we shall see how these are removed in the context of $R_{\rm h}=ct$,
though this would not be possible in $\Lambda$CDM. We begin in \S~II with a brief status
report on the $R_{\rm h}=ct$ model, and point to the various publications where
its predictions have been tested against the data. In \S~III, we discuss some
relevant observational issues pertaining to the CMB, including the interpretation
of the acoustic horizon as the characteristic length extracted from its power
spectrum. In \S~IV we describe the BAO scale and compare it to the acoustic
horizon in \S~V. In this section, we also discuss why the LSS had to be at
$z_{\rm cmb}\sim 16$ if these two scales are equal. In \S~VI we describe how
the CMB could have originated from dust opacity in this model, and we end
with an assessment of our results in \S\S~VII and VIII.

\section{The $R_{\rm h}=ct$ Model}
The $R_{\rm h}=ct$ universe has been described extensively in the
literature and its predictions have been tested against many observations
at high and low redshifts. This cosmology has much in common with $\Lambda$CDM,
but includes an additional ingredient motivated by several theoretical and
observational arguments \citep{Melia2007,Melia2016,Melia2017b,MeliaAbdelqader2009,MeliaShevchuk2012}.
Like $\Lambda$CDM, it also adopts the equation of state
$p=w\rho$, with $p=p_{\rm m}+ p_{\rm r}+p_{\rm de}$ and $\rho=\rho_{\rm m}+
\rho_{\rm r}+\rho_{\rm de}$, but goes one step further by specifying that
$w=(\rho_{\rm r}/3+ w_{\rm de}\rho_{\rm de})/\rho=-1/3$ at all times.
In spite of the fact that this prescription appears to be very different from
the equation of state in $\Lambda$CDM, where $w=(\rho_{\rm r}/3-\rho_\Lambda)/\rho$,
nature is in fact telling us that if we ignore the constraint $w=-1/3$ and
instead proceed to optimize the parameters in $\Lambda$CDM by fitting
the data, the resultant value of $w$ averaged over a Hubble time, is
actually $-1/3$ within the measurement errors. Thus, although
$w=(\rho_{\rm r}/3-\rho_\Lambda)/\rho$ in $\Lambda$CDM cannot be equal
to $-1/3$ from one moment to the next, its value averaged over the age
of the Universe is equal to what it would have been in $R_{\rm h}=ct$
all along.

This result does not prove that $\Lambda$CDM is incomplete,
but nonetheless suggests that the inclusion of the additional constraint
$w=-1/3$ might render its predictions closer to the data. By now, one-on-one
comparisons between $\Lambda$CDM and $R_{\rm h}=ct$ have been carried out
for a broad range of observations, from the angular correlation function
of the CMB \citep{MeliaGenova2018,Melia2014b} and high-$z$ quasars
\citep{Melia2013a,Melia2014c} in the early Universe, to gamma-ray bursts
\citep{Wei2013} and cosmic chronometers \citep{MeliaMaier2013} at
intermediate redshifts and, most recently, to the relatively nearby Type
Ia SNe \citep{Wei2015}. The application of model selection tools to
these tests indicates that the likelihood of $R_{\rm h}=ct$ being `closer
to the correct model' is typically $\sim 90\%$ compared to only
$\sim 10\%$ for $\Lambda$CDM. And most recently, the Alcock-Pacz\'ynski
test using BAO measurements has been shown to favour $R_{\rm h}=ct$
over $\Lambda$CDM at high redshifts \citep{MeliaLopez2017}.

There is therefore ample reason to consider the viability of the
$R_{\rm h}=ct$ Universe, and to see how one might interpret the formation
of the CMB in this model. This is one of several remaining critical
tests facing the $R_{\rm h}=ct$ universe. We recently demonstrated that,
while the angular correlation function of the CMB as measured with the
latest {\it Planck} release \citep{Planck2016a} remains in tension
with the predictions of $\Lambda$CDM, it is consistent with $R_{\rm h}=ct$
\citep{MeliaGenova2018}. It is still not clear, however,
whether the power spectrum itself may be fully explained in this model.
This paper is an important step in that direction. A second issue is whether
big bang nucleosynthesis is consistent with the constant expansion rate
required in this cosmology. It has been known for several decades that
a linear expansion with the physical conditions in the early $\Lambda$CDM
universe simply doesn't work because the radiation temperature and densities
don't scale properly with redshift \citep{Kaplinghat2000,Sethi2005}.
In $R_{\rm h}=ct$, however, the total equation of state is the
zero active mass condition $\rho+3p=0$, in terms of the total energy
density and pressure, so the various constituents in the cosmic fluid
evolve differently than those in the standard model. The situation is
closer to the so-called Dirac-Milne universe \citep{Benoit2012},
which also has linear expansion, so the outlook is more promising.

In reality, the standard model has not yet completely solved big
bang nucleosynthesis. The yields are generally consistent with the
observed abundances for $^4$He, $^3$He, and $D$, but $^7$Li
is over-produced by a significant amount \citep{Cyburt2008}.
This problem will go away with $R_{\rm h}=ct$ nucleosynthesis,
which is a two-step process, first through the thermal and homogeneous
production of $^4$He and $^7$Li, and then via the production of $D$ and
$^3$He. Previous work, e.g., by \cite{Benoit2012},
suggests that the timeline in this model is greatly different from that
in $\Lambda$CDM. Whereas all of the burning must take place before
neutrons decay in the latter, nucleosynthesis is a much slower process
in the former, with a neutron pool sustained via weak interactions.
The burning rate is much lower, but its duration is significantly longer,
so the $^4$He is produced over a hundred million years instead of only
15 minutes. According to these earlier simulations, the
Lithium anomaly largely disappears because the physical conditions
during the nuclear burning are far less extreme than in $\Lambda$CDM.
This work is well outside the scope of the present paper, of course,
but we highlight it here as one of the principal remaining problems
to address with this new cosmology.

The various measures of distance and time in the $R_{\rm h}=ct$
universe take on very simple forms, with very few parameters
\citep{Melia2007,MeliaShevchuk2012,MeliaMaier2013}. In some applications,
there are no parameters at all, making the analysis very straightforward,
and the results relatively unambiguous. For example,
the Hubble constant is $H(z) = H_0(1+z)$,
and the age is $t=1/H$, so the age-redshift relationship is
\begin{equation}
t(z) = {1\over H_0(1+z)}\;.
\end{equation}
And since $a(t)=(t_0/t)$ in this cosmology, we also have
\begin{equation}
(1+z)={a(t)\over a(t_0)}={t_0\over t}\;.
\end{equation}

Given the constraint on density and pressure alluded to above,
it is not difficult to show how the energy density of the various constituents
must evolve with redshift in this cosmology. Putting
\begin{equation}
\rho=\rho_{\rm r}+\rho_{\rm m}+\rho_{\rm de}\;,
\end{equation}
and
\begin{equation}
p=-\rho/3=w_{\rm de}\rho_{\rm de}+\rho_{\rm r}/3\;,
\end{equation}
we immediately see that
\begin{equation}
\rho_{\rm r}=-3w_{\rm de}\rho_{\rm de}-\rho\;,
\end{equation}
under the assumption that $p_{\rm r}=\rho_{\rm r}/3$ and
$p_{\rm m}\approx 0$. Throughout the cosmic evolution,
\begin{equation}
\rho(t)=\rho_{\rm c}\,a(t)^{-2}\;,
\end{equation}
where $\rho_{\rm c}\equiv 3c^2H_0^2/8\pi\,G$ is the critical density and
$a(t_0)=1$ in a flat universe.

\begin{figure}[h]
\vskip 0.2in
\includegraphics[width=1.0\linewidth]{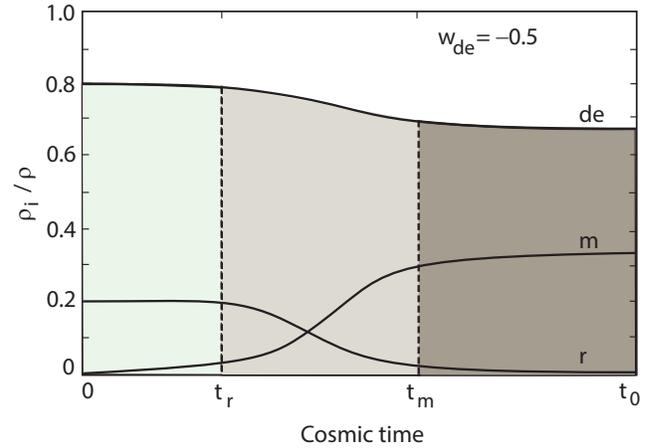}
\caption{Schematic diagram illustrating a possible evolution of
the various constituents $\rho_i$---dark energy (de), radiation (r) and
matter (m)---in $R_{\rm h}=ct$, as a function of cosmic time.
The conditions today imply that $w_{\rm de}=-0.5$, which then
fixes $\rho_{\rm r}/\rho=0.2$ and $\rho_{\rm de}/\rho=0.8$
at $z\gg 1$, while $\rho_{\rm m}/\rho=1/3$ and $\rho_{\rm de}/
\rho=2/3$ for $z\sim 0$. Radiation is dominant over matter in
the region $t< t_{\rm r}$, while matter dominates over
radiation for $t>t_{\rm m}$.}
\end{figure}

Equation~(5) constrains the radiation energy density in terms
of dark energy and $\rho$ at any epoch. At low redshifts, however,
we also know that the CMB temperature ($T_0\approx 2.728$ K) translates
into a normalized radiation energy density $\Omega_{\rm r}\approx
5\times 10^{-5}$, which is negligible compared to matter
and dark energy. Throughout this paper, the mass fractions
$\Omega_{\rm m} \equiv \rho_{\rm m}/\rho_{\rm c}$, $\Omega_{\rm r}\equiv
\rho_{\rm r}/\rho_{\rm c}$, and $\Omega_{\rm de}\equiv \rho_{\rm de}/
\rho_{\rm c}$, are defined in terms of the current matter ($\rho_{\rm m}$),
radiation ($\rho_{\rm r}$), and dark energy ($\rho_{\rm de}$) densities,
and the critical density $\rho_{\rm c}$.
Therefore, $w_{\rm de}$ must be $\sim -1/2$ in order to
produce a partitioning of the constituents in line with what we see in the
local Universe. With this value,
\begin{equation}
\Omega_{\rm de}= -{1\over 3w_{\rm de}}= {2\over 3}\;,
\end{equation}
while
\begin{equation}
\Omega_{\rm m}= {1+3w_{\rm de}\over 3w_{\rm de}}= {1\over 3}
\end{equation}
where, of course, $\Omega_{\rm m}=\Omega_{\rm b}+\Omega_{\rm d}$, representing
both baryonic and dark matter \citep{MeliaFatuzzo2016}.

At the other extreme, when $z\gg 1$, it is reasonable to hypothesize
that $\rho$ is dominated by radiation and dark energy\footnote{In the 
context of $R_{\rm h}=ct$, we know that radiation alone cannot sustain an 
equation of state $p=-\rho/3$, so dark energy is a necessary ingredient.}, so that
$\rho\approx \rho_{\rm r}+\rho_{\rm de}$. In that case, one would have
\begin{equation}
\rho_{\rm de}\approx {2\over 1-3w_{\rm de}}\rho_{\rm c}(1+z)^2\quad (z\gg 1)\;,
\end{equation}
and
\begin{equation}
\rho_{\rm r}\approx {3w_{\rm de}+1\over 3w_{\rm de}-1}\rho_{\rm c}(1+z)^2\quad (z\gg 1)\;,
\end{equation}
implying a relative partitioning of $\rho_{\rm de}=0.8\rho$ and
$\rho_{\rm r}= 0.2\rho$ (if $w_{\rm de}$ continues to be constant
at $-1/2$ towards higher redshifts). In other words, the zero active mass
condition $\rho+3p=0$ would be consistent with a gradual transition
of the equilibrium representation of the various constituents from the
very early universe, in which $\rho_{\rm de}/\rho=0.8$, to the
present, where $\rho_{\rm de}/\rho=2/3$. And during this evolution,
the radiation energy density that is dominant at $z\gg 1$, with
$\rho_{\rm r}/\rho= 0.2$, would eventually have given way to matter with
$\rho_{\rm m}/\rho= 1/3$ at later times ($z\sim 0$). This evolution
is shown schematically in figure~1. As we shall see shortly, the
physical properties of the medium at the LSS---presumably falling
between $t_{\rm r}$ and $t_{\rm m}$---provide a valuable datum in
between these two extreme limits (i.e., $t_{\rm r}\lesssim t
\lesssim t_{\rm m}$).

\begin{figure}[h]
\vskip 0.2in
\includegraphics[width=1.0\linewidth]{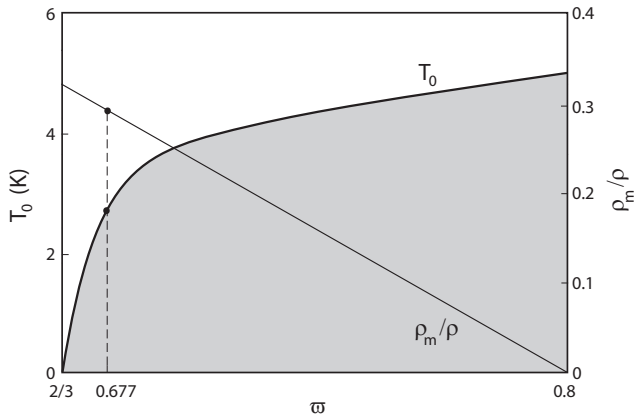}
\caption{The CMB temperature $T_0$ today, as a function of the fractional
representation of dark energy, $\varpi\equiv \rho_{\rm de}/\rho$, at the redshift
$z_{\rm cmb}$ of the last scattering surface. Also shown is the fractional representation
of matter, $\rho_{\rm m}/\rho$, at $z_{\rm cmb}\approx 16$. The Universe is
completely dominated by dark energy and radiation ($\varpi=0.8$) at $z\gg 1$, and
by dark energy and matter ($\varpi=2/3$) at low redshifts. Quite remarkably,
$T_0\lesssim 5$ K for all values of $\varpi$, but matches the specific measured
temperature $2.728$ K when $\varpi=0.677$, at which point one also finds
a matter representation $\rho_{\rm m}/\rho=0.308$.}
\end{figure}

Let us now define the ratio $\varpi\equiv \rho_{\rm de}/\rho$.
On the basis of the two arguments we have just made, we expect that
$0.8\ge\varpi\ge 2/3$ throughout the history of the Universe. Solving
Equations~(3) and (4), we therefore see that, at any redshift,
\begin{equation}
\rho_{\rm r}=\left({3\over 2}\varpi-1\right)(1+z)^2\rho_{\rm c}\;,
\end{equation}
while
\begin{equation}
\rho_{\rm m}=\left(2-{5\over 2}\varpi\right)(1+z)^2\rho_{\rm c}\;.
\end{equation}
Of course, the fact that $\rho_{\rm r}$ is constrained by the expression
in Equation~(10) at large redshift means that the radiation
is coupled to dark energy in ways yet to be determined through the
development of new physics beyond the standard model. Nonetheless,
for specificity, we will also assume that the radiation is always
a blackbody, both at high and low redshifts, though with one important
difference---that the relic photons are freely streaming below the
redshift $z_{\rm cmb}$ at the last scattering surface, corresponding to a time
$t_{\rm r}<t_{\rm cmb}<t_{\rm m}$ in figure~1, at which the radiation effectively
`decouples" from the other constituents. Therefore
\begin{equation}
T(z) = T_0(1+z)\quad (z\lesssim z_{\rm cmb})\;.
\end{equation}

At very high redshifts, however, $T$ is given explicitly by the redshift
dependence of $\rho_{\rm r}$. We still do not know precisely where the radiation
decouples from matter and dark energy, and begins to stream freely according
to the expression in Equation~(13) but, as we shall see below, our results are
not strongly dependent on this transition redshift, principally because $\varpi$ is
so narrowly constrained to the range $(2/3,0.8)$. Thus, for simplicity, we shall
assume that for $z>z_{\rm cmb}$ we may
put\footnote{In this expression, we have adopted the {\it Planck}
optimized value of the Hubble constant, $H_0=67.6\pm0.9$ km s$^{-1}$
Mpc$^{-1}$ (Planck Collaboration 2014). To be fair, this is the value measured in
the context of $\Lambda$CDM, and while a re-analysis of the {\it Planck}
data in the context of $R_{\rm h}=ct$ will produce a somewhat different
result for $H_0$, the differences are likely to be too small to affect
the discussion in this paper.}
\begin{equation}
T(z) \approx 31.8\;{\rm K}\;(3\varpi/2-1)^{1/4}(1+z)^{1/2} \quad (z\gtrsim z_{\rm cmb})\;.
\end{equation}
Even before considering the consequences of identifying the BAO scale
as the acoustic horizon, which we do in the next section, we can already
estimate the location of the LSS by setting Equation~(13) equal to (14),
which yields
\begin{equation}
T_0\approx 8.53\,\left(\varpi(z_{\rm cmb})-{2\over 3}\right)^{1/4}\;{\rm K}\;.
\end{equation}
Remembering that $0.8\ge\varpi\ge2/3$ everywhere, we therefore see
that $T_0$ in this model must be $\lesssim 5$ K, no matter where the
LSS is located. This is quite a remarkable result because the only
input used to reach this conclusion is the value of $H_0$, unlike
the situation with $\Lambda$CDM, in which one must assume
both a value of $H_0$ and optimize the baryon to photon
fraction in the early Universe to ensure a value of $T_0$ in this range.
Figure~2 illustrates how $T_0$ today changes with $\varpi$ if
we assume $z_{\rm cmb}=16$ (see below). We see that $\varpi(z_{\rm cmb})$
must then be $\approx 0.677$ when we fix $T_0=2.728$ K, which is
consistent with $t_{\rm cmb}$ being closer to $t_{\rm m}$ than $t_{\rm r}$
in figure~1. Indeed, we find from Equations~(11) and (14) that, at $z=z_{\rm cmb}$,
$\rho_{\rm r}/\rho \sim 0.016$ and $\rho_{\rm m}/\rho\sim 0.308$.

We shall consider the more specific constraints imposed by the
CMB acoustic horizon and the BAO peak measurements shortly, but for
now we have already demonstrated a very powerful property of the
$R_{\rm h}=ct$ universe---that $H_0$ and the baryon
to photon ratio are not independent of each other. And
clearly, while $z_{\rm cmb}\sim 1080$ in $\Lambda$CDM, the LSS
must occur at a much lower redshift ($z_{\rm cmb}\lesssim 30$) in
this model.

The temperature calculated from Equations~(13) and (14) is compared to
that of the standard model in figure~3. This figure also indicates
the location of $z_{\rm cmb}$ based on the argument in the previous
paragraph, which will be bolstered shortly with constraints from the
acoustic and BAO scales. Thus, while $T\sim 3,000$ K at $z\sim 1080$ in
$\Lambda$CDM, so that hydrogen `recombination' may be relevant to the CMB
in this model, the temperature is too low at $z_{\rm cmb}< 30$ for
this mechanism to be responsible for liberating the relic photons
in $R_{\rm h}=ct$.

\begin{figure}[h]
\vskip 0.2in
\includegraphics[width=1.0\linewidth]{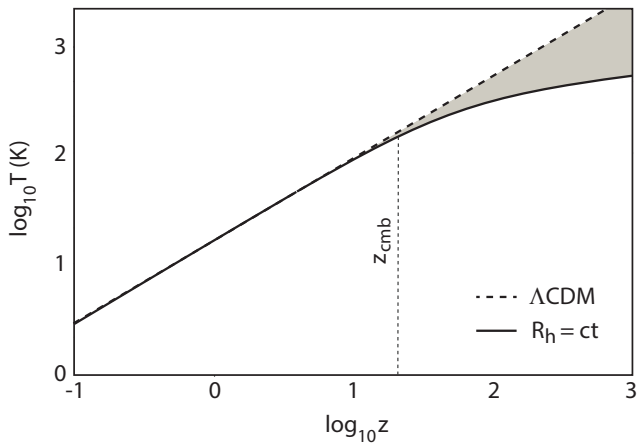}
\caption{The CMB temperature in $R_{\rm h}=ct$ (solid) compared to
its counterpart in $\Lambda$CDM (dashed). The location of the LSS
at $z_{\rm cmb}\sim 16$ in the former model is based on several
observational arguments (see text). By comparison, $z_{\rm cmb}\sim
1080$ in the standard model.}
\end{figure}

In this regard, our reconsideration of dust's
contribution to the formation of the CMB deviates from the original proposal
\citep{Rees1978}, in that the radiation being rethermalized at $z\sim
z_{\rm cmb}$ in this picture need not all have been emitted by Pop III stars.
Indeed, given that $\rho_{\rm r}\approx 0.2\rho$ for $z\gg 1$, these photons
were more likely produced during the intervening period between the big bang
and decoupling prior to the reprocessing by dust at $z\sim 16$. The implied
coupling between radiation and the rest of the cosmic fluid at high redshifts
requires physics beyond the standard model, which acted to maintain the
$\sim 0.2\rho$ fraction until decoupling, after which the radiation
streamed freely---except at $z\sim 15-20$, where it would have attained
thermal equilibrium with the dust. An important caveat with this procedure
is that we are ignoring the possible role played by other relativistic species,
whose presence would affect the redshift dependence of the temperature
$T$. Certainly the early presence of energetic neutrinos may have affected
structure formation in $\Lambda$CDM. But given that we know very little
about extensions to the standard model, we shall for simplicity assume
that such particles will not qualitatively impact $T(z)$, though recognize
that this assumption may have to be modified, or supplanted, when more
is known. This caveat notwithstanding, the dust in this picture would have
had no influence on the value of $\rho_{\rm r}$, but simply reprocessed all
components (if more than one) in the radiation field into the single,
blackbody CMB we see today.

We can say with a fair degree of certainty, however, that---as in the
standard model---the background radiation field would not have
been significantly influenced by Pop III star formation. We shall demonstrate
in \S~VI.2 below that more recent work has shown that the halo
abundance was probably orders of magnitude smaller than previously thought
\citep{Johnson2013}, greatly reducing the likely contribution ($\lesssim 0.5\%$)
of Pop III stars to the overall radiative content of the Universe at that time.
Thus, the original proposal by Rees \citep{Rees1978} and others would not work because
Pop III stars could not supply more than this small fraction of the photons
that were thermalized by the dust they ejected into the interstellar medium.

In \S~VI below, we will consider three of the most important diagnostics
regarding whether or not the CMB and its fluctuations (at a level of 1 part
per 100,000) were produced at recombination, or much later by dust emission
at the transition from Pop III to Pop II stars (i.e., $z_{\rm cmb}< 30$). An
equally important feature of the microwave temperature is its isotropy across
the sky. Inflation ensures isotropy in $\Lambda$CDM, but what about
$R_{\rm h}=ct$?  This question is related to the broader horizon problem,
which necessitated the creation of an inflationary paradigm in the first
place. It turns out, however, that the horizon problem is an issue only for
cosmologies that have a decelerated expansion at early times. For a constant
or accelerated expansion, as we have in $R_{\rm h}=ct$, all parts of
the observable universe today have been in equilibrium from the earliest
moments \citep{Melia2013b}. Thus, not only has everything in the observable
($R_{\rm h}=ct$) universe been homogeneous from the beginning, it has
also been distributed
isotropically as well. This includes the energy density $\rho$ and its
fluctuations, the Pop III stars that formed from them under the action of
self gravity, and the dust they expelled into the interstellar medium prior to
the formation of large-scale structure. And since the radiative energy density
$\rho_{\rm r}$ and its temperature (see Eq.~11) were also distributed
homogeneously and isotropically prior to rethermalization by dust,
the eventual CMB produced at $z_{\rm cmb}< 30$, and its tiny fluctuations,
would therefore now also be isotropic across the sky. In other words, an
isotropic CMB cannot be used to distinguish between $R_{\rm }=ct$ and the
inflationary $\Lambda$CDM.

\section{The Acoustic Scale}
CMB experiments, most recently with {\it Planck} \citep{Planck2014},
have identified a scale $r_{\rm s}$ in both the temperature and polarization
power spectrum, with a measured angular size $\theta_{\rm s}=(0.596724\pm
0.00038)^\circ$ on the LSS. If this is an acoustic horizon, the CMB fluctuations
have a characteristic size $\theta_{\rm f}\approx 2\theta_{\rm s}$, since the
sound wave produced by the dark-matter condensation presumably expanded as a
spherical shell and what we see on the LSS is a cross section of this structure,
extending across twice the acoustic horizon. Since the multipole number is
defined as $l_{\rm s}=2\pi/\theta_{\rm f}$, one has $l_{\rm s}=\pi/\theta_{\rm s}$,
which produces the well-known location (at $\sim 300$) of the first peak with an
acoustic angular size $\theta_{\rm s}\sim 0.6^\circ$. Actually, there are several
additional physical effects one must take into account in order to arrive at the
true measured value of $l^{TT}_m$ for the first peak. These include the decay of
the gravitational potential and contributions from the Doppler shift of the
oscillating fluid, all of which introduce a phase shift $\phi_m$ in the spectrum
\citep{Doran2002,Page2003}. The general relation for all peaks and
troughs is $l^{TT}_m=l_{\rm s}(m-\phi_m)$. Thus, since $\phi_m$ is typically
$\sim 25\%$, the measured location of the first peak ends up at $l^{TT}_1\sim 220$.

The acoustic scale in any cosmological model depends critically on when
matter and radiation decoupled ($t_{\rm dec}$) and how the sound speed $c_{\rm s}$
evolved with redshift prior to that time. In $\Lambda$CDM, the decoupling
was completed at recombination. But this need not be the case in every model.
As we shall see, the radiation may have decoupled from matter earlier than the time at which
the observed CMB was produced if, as in the case of $R_{\rm h}=ct$, rethermalization
of the photons by dust occurred at $z<30$. For the rest of this paper, we therefore
make a distinction between $t_{\rm dec}$ and $t_{\rm cmb}$.

Since Hydrogen was the
dominant element by number, the transition from an optically thick to thin medium is
thought to have occurred when the number of ambient H-ionizing photons dropped
sufficiently for Hydrogen to recombine \citep{Peebles1970,Hu1995,White1994}.
The actual estimate of the rate at which neutral Hydrogen
formed also depends on other factors, however, since the $13.6$ eV photons couldn't
really `escape' from the fluid. Instead, the process that took photons out of the
loop was the $2s\rightarrow 1s$ transition, which proceeds via 2-photon emission
to conserve angular momentum. So neutral Hydrogen did not form instantly; the
epoch of recombination is thought to have coincided with the fraction $x$ of
electrons to baryons dropping below $50\%$. But because the baryon to photon
ratio is believed to have been very small (of order $10^{-9}$ in some
models), the H-ionizing photons did not have to come from the center
of the Planck distribution. There were enough ionizing photons in the
Wien tail to ionize all of the Hydrogen atoms. This disparity in number
means that the value of the radiation temperature at decoupling is
poorly constrained, in the sense that $x$ would have depended on the
baryon to photon ratio as well as temperature. But since the dependence
of $x$ on the baryon density $\rho_{\rm b}$ was relatively small compared
to its strong exponential dependence on temperature, any model change in
$\rho_{\rm b}$ could easily have been offset by a very tiny change in
temperature. So $z_{\rm dec}$ is nearly independent of the global
cosmological parameters, and is determined principally by the
{\it choice} of $r_{\rm s}$, which is typically calculated according to
\begin{equation}
r_{\rm s}\equiv \int_0^{t_{\rm dec}} c_{\rm s}(t^{\prime})[1+z(t^{\prime})]\,dt^\prime\;,
\end{equation}
from which one then infers a proper distance
$R_{\rm s}(z_{\rm dec})=r_{\rm s}/(1+z_{\rm dec})$ traveled by the sound wave
reaching the redshift at decoupling.

For a careful determination of $z_{\rm dec}$, one therefore needs to know how
the sound speed $c_{\rm s}$ evolves with time. For a relativistic fluid,
$c_{\rm s}=c/\sqrt{3}$, but the early universe contained matter as well
as radiation, and dark energy in the context of $R_{\rm h}=ct$. And though
the strong coupling between photons, electrons and baryons allows us to
treat the plasma as a single fluid for dynamical purposes during this era
\citep{Peebles1970}, the contribution of baryons to the equation of state
alters the dependence of $c_{\rm s}$ on redshift, albeit by a modest amount.
For example, a careful treatment of this quantity in the context of
$\Lambda$CDM takes into account its evolution with time, showing that
differences amounting to a factor $\sim 1.3$ could lead to a reduction
in sound speed. Quantitatively, such effects are typically rendered
through the expression
\begin{equation}
c_{\rm s}={c\over\sqrt{3(1+3\rho_{\rm b}/4\rho_{\rm r})}}
\end{equation}
\citep{White1994}. Obviously, $c_{\rm s}$ reduces to $c/\sqrt{3}$ when
$\rho_{\rm b}/\rho_{\rm r}\rightarrow 0$, as expected.

The situation in $R_{\rm h}=ct$ is somewhat more complicated, primarily
because $\rho$ contains dark energy throughout the cosmic expansion.
From \S~II, we expect that $\rho_{\rm r}/\rho_{\rm m}$ is a decreasing
function of $t$. In addition, $\rho_{\rm r}$ is itself always a small
fraction of $\rho$, but in order to maintain the constant equation of
state $p=-\rho/3$, it is reasonable to expect that all three constituents
remain coupled during the acoustically important epoch, i.e., in the
region $t\lesssim t_{\rm m}$ in figure~1. Therefore,
\begin{equation}
c_{\rm s}^2=\left(+{1\over 3}\right){\partial\rho_{\rm r}\over\partial\rho}+
{\partial p_{\rm de}\over\partial\rho_{\rm de}}{\partial \rho_{\rm de}\over
\partial\rho}\;,
\end{equation}
under the assumption that $p_{\rm m}\approx 0$ at all times. We already
know that $\partial\rho_{\rm r}/\partial\rho\le 0.2$. Thus, depending on
the sound speed of dark energy, the overall sound speed in the cosmic fluid,
$c_{\rm s}$, may or may not be much smaller than $c/\sqrt{3}$ in the early
$R_{\rm h}=ct$ universe.

We can estimate its value quantitatively by assuming for simplicity that
\begin{equation}
c_{\rm s}(t) = c_{\rm s}(t_*)\left({t_*\over t}\right)^\beta\;,
\end{equation}
where $t_*$ is the time at which the acoustic wave is produced and
the index $\beta$ is positive in order to reflect the decreasing
importance of radiation with time. The acoustic radius in such a model
would therefore be given by the expression
\begin{equation}
r_{\rm s}^{R_{\rm h}=ct}(t_{\rm dec})=c_{\rm s}(t_*)\,t_0\,t_*^\beta\int_{t_*}^{t_{\rm dec}}
{dt^\prime\over (t^\prime)^{1+\beta}}\;.
\end{equation}
Thus, as long as $t_{\rm dec}\gg t_*$,
\begin{equation}
r_{\rm s}^{R_{\rm h}=ct}={c_{\rm s}(t_*)\,t_0\over\beta}
={R_{\rm h}(t_0)\over\beta}\left({c_{\rm s}(t_*)\over c}\right)\;,
\end{equation}
so that
\begin{equation}
\left({c_{\rm s}(t_*)\over c}\right)=\beta\left({r_{\rm s}^{R_{\rm h}=ct}\over R_{\rm h}(t_0)}\right)\;.
\end{equation}

We shall return to this after we discuss the BAO scale in the next section. Before doing so, however,
it is worthwhile reiterating an important difference between the acoustic scale in $\Lambda$CDM and
that in $R_{\rm h}=ct$. The consensus today is that, in the standard model, the temperature
of the baryon-photon fluid remained
high enough all the way to $t_{\rm cmb}$ for the plasma to be at least partially ionized, allowing a strong
coupling between the baryons and the radiation. As such, the comoving acoustic horizon $r_{\rm s}$
in Equation~(16) is calculated assuming that sound waves propagated continuously from $t\sim 0$
to $t_{\rm dec}\sim t_{\rm cmb}$. As one may see from Equation~(21), however, there are several
reasons why the analogous quantity $r_{\rm s}^{R_{\rm h}=ct}$ in $R_{\rm h}=ct$ may need to
be calculated with a truncated integral that does not extend all the way to $t_{\rm cmb}$. The
principal argument for this is that the kinetic temperature of the medium may have dropped below the ionization
level prior to the time at which the observed CMB was produced, which would effectively decouple
the baryons from the photons. This would certainly occur if rethermalization of the primordial radiation
field by dust happened at $z<30$, well after decoupling. Nonetheless, none of the analysis carried
out in this paper is affected by this. All we need to assume is that the acoustic horizon at the
last scattering surface remained constant thereafter, including at the redshift where the
BAO peaks are observed. To be clear, the physical scale $R_{\rm s}(t)=a(t)r_{\rm s}$ of the BAO
peaks is larger than that at $z_{\rm cmb}$, but this change is due solely to the effects of expansion,
arising from the expansion factor $a(t)$, not to a continued change in the comoving scale $r_{\rm s}$.
Thus, our imprecise knowledge of  the scale factor $r_{\rm s}^{R_{\rm h}=ct}$ in $R_{\rm h}=ct$
is not going to be an impediment to the analysis we shall be carrying out in this paper.

\section{The BAO Scale}
In tandem with the scale $\theta_{\rm s}$ seen by {\it Planck} and its
predecessors, a peak has also been seen in the correlation function of
galaxies and the Ly-$\alpha$ forest (see, e.g., \citealt{MeliaLopez2017},
and references cited therein). Nonlinear effects in
the matter density field are still mild at the scale where BAO would emerge,
so systematic effects are probably small and can be modeled with a low-order
perturbation theory \citep{Meiksin1999,Seo2005,Jeong2006,Crocce2006,Eisenstein2007b,Nishimichi2007,Matsubara2008,Padmanabhan2009,Taruya2009,Seo2010}.
Thus, the peak seen with large galaxy surveys can also be
interpreted in terms of the acoustic scale.

To be clear, we will be making the standard assumption that once the
acoustic horizon has been reached at decoupling, this scale remains fixed
thereafter in the comoving frame. The BAO proper scale, however, is not
the same as the acoustic proper scale in the CMB. Although these lengths
are assumed to be identical in the comoving frame, the horizon scale
continues to expand along with the rest of the Universe, according to
the expansion factor $a(t)$. As such, the physical BAO scale is actually
much bigger than the CMB acoustic length, with a difference that
depends critically on the cosmological model. As we shall see, this
is the reason the recombination picture does not work in $R_{\rm h}=ct$,
because equating these two scales in this model implies a redshift for
the CMB much smaller than 1080.

In the past several years, the use of reconstruction techniques
\citep{Eisenstein2007a,Padmanabhan2012} that enhance the quality of the galaxy
two-point correlation function and the more precise determination of the
Ly-$\alpha$ and quasar auto- and cross-correlation functions, has resulted
in the measurement of BAO peak positions to better than $\sim 4\%$ accuracy.
The three most significant of these are a) the measurement of the BAO peak
position in the anisotropic distribution of SDSS-III/BOSS DR12 galaxies
\citep{Alam2016} at the two independent/non-overlapping bins with $\langle
z\rangle=0.38$ and $\langle z\rangle=0.61$, using a technique of reconstruction
to improve the signal/noise ratio. Since this technique affects the position
of the BAO peak only negligibly, the measured parameters are independent
of any cosmological model; and b) the self-correlation of the BAO peak
in the Ly-$\alpha$ forest in the SDSS-III/BOSS DR11 data \citep{Delubac2015}
at $\langle z\rangle=2.34$, in addition to the cross-correlation of the
BAO peak of QSOs and the Ly-$\alpha$ forest in the same survey
\citep{Font2014}.

In their analysis of these recent measurements, \citep{Alam2016}
traced the evolution of the BAO scale separately over nearby redshift
bins centered at 0.38, 0.51 and 0.61 (the $z=0.51$ measurement is
included for this discussion, though its bin overlaps with both of
the other two), and then in conjunction with the Ly-$\alpha$ forest
measurement at $z=2.34$ \citep{Delubac2015}. As was the case in
\cite{MeliaLopez2017}, these authors opted not to include
other BAO measurements, notably those based on photometric clustering
and from the WiggleZ survey \citep{Blake2011}, whose larger
errors restrict their usefulness in improving the result. Older
applications of the galaxy two-point correlation function to measure
a BAO length were limited by the need to disentangle the acoustic
length in redshift space from redshift space distortions arising
from internal gravitational effects \citep{Lopez2014}. To do this,
however, one invariably had to either assume prior parameter values or
pre-assume a particular model to determine the degree of contamination,
resulting in errors typically of order $20-30\%$.

Even so, several inconsistencies were noted between theory and
observations at various levels of statistical significance. For example,
based on the BAO interpretation of a peak at $z=0.54$, the implied angular
diameter distance was found to be $1.4\sigma$ higher than what is expected
in the concordance $\Lambda$CDM model \citep{Seo2010}. When combined with
the other BAO measurements from SDSS DR7 spectroscopic surveys \citep{Percival2010}
and WiggleZ \citep{Blake2011}, there appeared to be a
tendency of cosmic distances measured using BAO to be noticeably larger
than those predicted by the concordance $\Lambda$CDM model.

The more recent measurements using several innovative reconstruction
techniques have enhanced the quality of the galaxy two-point correlation
function and the quasar and Ly-$\alpha$ auto- and cross-correlation
functions.  Unfortunately, in spite of this improved accuracy, the
comparison with model predictions depends on how one chooses the
data. When the Ly-$\alpha$ measurement at $z=2.34$ is excluded,
\cite{Alam2016} find that the BOSS measurements are fully
consistent with the {\it Planck} $\Lambda$CDM model results, with
only one minor level of tension having to do with the inferred
growth rate $f\sigma_8$, for which the BOSS BAO measurements
require a bulk shift of $\sim 6\%$ relative to {\it Planck}
$\Lambda$CDM. In all other respects, the standard model predictions
from {\it Planck} fit the BAO-based distance observables at these
three redshift bins typically within $1\sigma$.

On the other hand, \cite{Alam2016} also find that when the
Ly-$\alpha$ measurement at $z=2.34$ is included with the three lower
redshift BOSS measurements, the combined data deviate from the
concordance model predictions at a $2-2.5\sigma$ level. This result
has been discussed extensively in the literature
\citep{Delubac2015,Font2014,Sahni2014,Aubourg2015},
and is consistent with our previous analysis using a similar data set
to carry out an Alcock-Paczy\'nski (AP) test of various cosmological
models \citep{MeliaLopez2017}.

The AP test, based on the combined BOSS and Ly-$\alpha$
measurements (see Table 1 below), shows that the observations
are discrepant at a statistical significance of $\gtrsim 2.3\sigma$
with respect to the predictions of a flat $\Lambda$CDM cosmological model
with the best-fit {\it Planck} parameters \citep{MeliaLopez2017}.
More so than any other observation of the acoustic scale to date,
the tension between the measurement at $\langle z\rangle=2.34$ and theory
is problematic because the observed ratio $d_A/d_H=1.229\pm 0.11$ is
obtained independently of any pre-assumed model, in terms of the
angular-diameter distance $d_A(z)$ and Hubble radius $d_H(z)\equiv c/H(z)$.

The bottom line is that BAO measurements may or may not be in tension
with {\it Planck} $\Lambda$CDM, largely dependent on which measurements
one chooses for the analysis. Certainly, the BAO measurement based on the
Ly-$\alpha$ forest requires different techniques than those used with the
galaxy samples, and no doubt is affected by systematics possibly different
from those associated with the latter. For instance, \cite{Delubac2015}
worry about possible observational biases when examining
the Ly-$\alpha$ forest. What is clear up to this point is that, given the rather
small range in BOSS redshifts (essentially $0.38<z< 0.61$) one may adequately
fit the distance observables with either {\it Planck} $\Lambda$CDM or
$R_{\rm h}=ct$. The factor separating these two models is primarily
the inclusion of the Ly-$\alpha$ measurements at $z=2.34$ which,
however, is a different kind of observation, and may be problematic
for various reasons.

Table~1 lists the three measurements used to carry out the Alcock-Paczy\'nski
test in order to establish whether or not the BAO scale $r_{\rm BAO}$ is a
true `standard ruler' \citep{MeliaLopez2017}. The ratio
\begin{equation}
{\cal D}(z)\equiv d_A(z)/d_H(z)
\end{equation}
(e.g., from the flux-correlation function of the Ly-$\alpha$ forest of
high-redshift quasars \citep{Delubac2015}) is independent of both $H_0$
and the presumed acoustic scale $r_{\rm BAO}$, thereby providing a very
clean test of the cosmology itself.

In $\Lambda$CDM, $d_A$ depends on several parameters, including the mass
fractions $\Omega_{\rm m}$, $\Omega_{\rm r}$, and $\Omega_{\rm de}$.
Assuming zero spatial curvature, so that $\Omega_{\rm m}+\Omega_{\rm r}
+\Omega_{\rm de}=1$, the angular-diameter distance at redshift $z$
is given by the expression
\begin{eqnarray}
d^{{\Lambda}\rm CDM}_A(z)&=&{c\over H_0}{1\over (1+z)}\int_{0}^{z}
\left[\Omega_{\rm m}(1+u)^3+\right.\nonumber\\
&\null&\left.\hskip-0.5in\Omega_{\rm r}(1+u)^4 +\Omega_{\rm de}
(1+u)^{3(1+w_{\rm de})}\right]^{-1/2}\,du\,,
\end{eqnarray}
where $p_{\rm de}=w_{\rm de}\rho_{\rm de}$ is the dark-energy equation
of state. Thus, since $\rho_{\rm r}$ is known from the CMB temperature
$T_0=2.728$~K today, the essential free parameters in flat $\Lambda$CDM
are $H_0$, $\Omega_{\rm m}$ and $w_{\rm de}$, though the scaled baryon
density $\Omega_{\rm b}\equiv \rho_{\rm b}/\rho_{\rm c}$ also enters
through the sound speed (Eq.~17). The other quantity in Equation~(23)
is the Hubble distance,
\begin{eqnarray}
d^{{\Lambda}\rm CDM}_{\rm H}(z) &\equiv&{c\over H(z)}\nonumber \\
&=& {c\over H_0}\left[\Omega_{\rm m}(1+z)^3+\Omega_{\rm r}(1+z)^4\right.\nonumber\\
&\null&\left.\qquad+\Omega_{\rm de} (1+z)^{3(1+w_{\rm de})}\right]^{-1/2}.
\end{eqnarray}
In the $R_{\rm h}=ct$ Universe, the angular-diameter distance is simply given as
\begin{equation}
d^{R_{\rm h}=ct}_A(z)=\frac{c}{H_{0}}\frac{1}{(1+z)}\ln(1+z)\;,
\end{equation}
while the Hubble distance is
\begin{equation}
d^{R_{\rm h}=ct}_{\rm H}(z)={c\over H_0}{1\over(1+z)}\;.
\end{equation}
In this cosmology, one therefore has the simple, elegant expression
\begin{equation}
{\cal D}_{R_{\rm h}=ct}(z)=\ln(1+z)\;,
\end{equation}
which is {\it completely free of any parameters}.

For $\Lambda$CDM with flatness as a prior, ${\cal D}_{\Lambda\rm CDM}$
relies entirely on the variables $\Omega_{\rm m}$ and
$w_{\rm de}$. This clear distinction between
${\cal D}_{\Lambda\rm CDM}(z)$ and ${\cal D}_{R_{\rm h}=ct}(z)$ can
therefore be used to test these competing models in a one-on-one comparison,
free of the ambiguities often attached to data tainted with nuisance parameters.
Unlike those cases, the measured ratio ${\cal D}_{\rm obs}$ is completely
independent of the model being examined. In \cite{MeliaLopez2017},
we used the Alcock-Paczy\'nski test to compare these model independent data
to the predictions of $\Lambda$CDM and $R_{\rm h}=ct$ and showed that the
standard model is disfavoured by these measurements at a significance greater
than $\sim 2.3\sigma$, while the probability of $R_{\rm h}=ct$ being consistent
with these observations is much closer to 1.

\begin{table*}
\vskip 0.2in
\center
{\footnotesize
\centerline{{\bf Table 1.} Inferred BAO scale, $r_{\rm BAO}$, from the most recent high-precision measurements}
\begin{tabular}{cccccc}
\\
\hline\hline
$z$\qquad&\qquad{${\cal D}_{\rm obs}(z)$}\qquad&\qquad$\theta_{\rm BAO}$\qquad&
\qquad{$r_{\rm BAO}^{\Lambda{\rm CDM}}$}\qquad &\qquad$r_{\rm BAO}^{R_{\rm h}=ct}$
&\qquad{\rm Reference}\qquad\\
\qquad&&\qquad{(deg)}\qquad&\qquad (Mpc)\qquad& \qquad{(Mpc)}\qquad &\qquad\qquad\\
\hline
0.38\qquad   & \qquad $0.286\pm0.025$    &       \qquad $5.60\pm 0.12$  &  \qquad $158.6\pm3.4$  &
\qquad $130.3\pm2.8$ & \qquad \cite{Alam2016}   \\
0.61\qquad   & \qquad $0.436\pm0.052$    &       \qquad $3.67\pm 0.08$  &  \qquad $153.7\pm3.4$  &
\qquad $126.3\pm2.8$ & \qquad \cite{Alam2016}   \\
2.34\qquad   & \qquad $1.229\pm0.110$    &       \qquad $1.57\pm 0.05$  &  \qquad $149.7\pm4.8$  &
\qquad $136.8\pm4.4$ & \qquad \cite{Delubac2015}  \\
\hline
{\rm Average}& & & \qquad $154.0\pm3.6$ & \qquad $131.1\pm4.3$ & \\
\hline\hline
\end{tabular}
}
\vskip 0.3in
\end{table*}

The inclusion of the BAO measurement at $z=2.34$ creates tension with the
$\Lambda$CDM interpretation of the acoustic scale, which is eliminated in
$R_{\rm h}=ct$, lending some support to the idea that the BAO and CMB acoustic
scales should be related in this model. For the application in this
paper, we must adopt a particular
value of $H_0$ to use these high-precision data to extract a comoving BAO
scale. For $\Lambda$CDM, we adopt the concordance parameter values
$\Omega_{\rm m}=0.31$, $H_0=67.6$ km s$^{-1}$ Mpc$^{-1}$, $w_{\rm de}=-1$,
and $\Omega_{\rm b}=0.022/h^2$ and, to keep the comparison as simple
as possible, we here assume the same value of $H_0$ for the $R_{\rm h}=ct$
cosmology. From the data in Table~1, we see that the scale $r_{\rm BAO}$ may be
used as a standard ruler over a significant redshift range ($0\le z\le 2.34$) in
both models, though the actual value of $r_{\rm BAO}$ is different if the same
Hubble constant is assumed in either case. Based solely on this outcome, the
interpretation of $r_{\rm BAO}$ as an acoustic scale could be valid in
$R_{\rm h}=ct$, perhaps more so than in $\Lambda$CDM.

\section{Adopting the Acoustic Horizon as a Standard Ruler}
Let us now assume that the BAO and CMB acoustic scales are equal. In the $R_{\rm h}=ct$
universe, we therefore have
\begin{equation}
\ln(1+z_{\rm cmb})={r_{\rm BAO}^{R_{\rm h}=ct}\over R_{\rm h}(t_0)\,\theta_{\rm s}}\;,
\end{equation}
so that
\begin{equation}
z_{\rm cmb}=16.05^{+2.4}_{-2.0}\;,
\end{equation}
which corresponds to a cosmic time $t_{\rm cmb}\approx 849$ Myr.
This redshift at last scattering in $R_{\rm h}=ct$ is quite different from
the corresponding value ($\sim 1080$) in $\Lambda$CDM, so is there any confirming
evidence to suggest that this is reasonable? There is indeed another
type of observation supporting this inferred redshift. The value
quoted in Equation~(30) is a good match to the $z_{\rm cmb}$ measured using
an entirely different analysis of the CMB spectrum, which we now describe.

It has been known for almost two decades that the lack of large-angle
correlations in the temperature fluctuations observed in the CMB is in
conflict with predictions of inflationary $\Lambda$CDM. Probabilities
($\lesssim 0.24\%$) for the missing correlations disfavour inflation at
better than $3\sigma$ \citep{Copi2015}. Recently, we \citep{MeliaGenova2018}
used the latest {\it Planck} data release \citep{Planck2014}
to demonstrate that the absence of large-angle correlations
is best explained with the introduction of a non-zero minimum wavenumber
$k_{\rm min}$ for the fluctuation power spectrum $P(k)$. This is an
important discriminant among different cosmological models because
inflation would have stretched all fluctuations beyond the horizon,
producing a $P(k)$ with $k_{\rm min}=0$ and, therefore, strong correlations
at all angles. A non-zero $k_{\rm min}$ would signal the presence of a
maximum fluctuation wavelength at decoupling, thereby favouring
non-inflationary models, such as $R_{\rm h}=ct$, which instead
produce a fluctuation spectrum with wavelengths no bigger than the
gravitational (or Hubble) radius \citep{MeliaGenova2018}.

It is beyond the scope of the present paper to discuss in
detail how the cutoff $k_{\rm min}$ impacts the role of inflation within
the standard model, but it may be helpful to place this measurement in
a more meaningful context by summarizing the key issue (see 
\citealt{Liu2020} for a more in-depth discussion). Slow-roll inflation 
in the standard model is viewed as the critical mechanism that can
simultaneously solve the horizon problem and generate a near scale-free 
fluctuation spectrum, $P(k)$. It is readily recognized that these two processes 
are intimately connected via the initiation of the inflationary phase, which in
turn also determines its duration. 

The identification of a cutoff $k_{\rm min}$ in $P(k)$ tightly constrains the
time at which inflation could have started, requiring the
often used small parameter $\epsilon$ \citep{Liddle1994} to be $\gtrsim 0.9$ 
throughout the phase of inflationary expansion in order to produce sufficient 
dilation to fix the horizon problem. Such high values of $\epsilon$ predict 
extremely red spectral indices, however, which disagree with measured near 
scale-free spectrum, which typically requires $\epsilon\ll 1$. Extensions to 
the basic picture have been suggested by several workers
\citep{Destri2008,Scacco2015,Santos2018,Handley2014,Ramirez2012,Remmen2014},
most often by adding a kinetic-dominated or radiation-dominated phase preceding 
the slow-roll expansion. But none of the approaches suggested thus far have been
able to simultaneously fix the horizon problem and produce enough expansion
to overcome the horizon problem. It appears that the existence of
$k_{\rm min}$ requires a modification and/or a replacement of the basic
inflationary picture \citep{Liu2020}.

In the $R_{\rm h}=ct$ cosmology, on the other hand, fluctuation modes never cross 
back and forth across the Hubble horizon, since the mode size and the Hubble radius 
grow at the same rate as the Universe expands. Thus, $k_{\rm min}$ corresponds to 
the first mode emerging out of the Planck domain into the semi-classical Universe 
\citep{Melia2019}. The scalar-field required for this has an exponential
potential, but it is not inflationary, and it satisfies the zero active mass
condition, $\rho_\phi+3p_\phi=0$, just like the rest of the Universe during
its expansion history. The amplitude of the temperature anisotropies observed
in the CMB requires the quantum fluctuations in $\phi$ to have classicalized
at $\sim 3.5\times 10^{15}$ GeV, suggesting an interesting physical connection
to the energy scale in grand unified theories. Indeed, such scalar-field 
potentials have been studied in the context of Kaluza-Klein cosmologies, string
theory and supergravity (see, e.g., \citealt{Halliwell1987}).

In terms of the variable
\begin{equation}
u_{\rm min}\equiv k_{\rm min}\,c\Delta\tau_{\rm cmb}\;,
\end{equation}
where $c\Delta\tau_{\rm cmb}$ is the comoving radius of the last scattering
surface written in terms of the conformal time difference between $t_0$ and
$t_{\rm cmb}$, the recent analysis of the CMB anisotropies \citep{MeliaGenova2018}
shows that the angular-correlation function anomaly disappears completely 
for $u_{\rm min}=4.34\pm 0.50$, a result that argues against the basic 
slow-roll inflationary paradigm for the origin and growth of perturbations 
in the early Universe, as we have just discussed. With an implied $u_{\rm min}=0$, 
the standard inflationary cosmology in its present form is disfavoured by 
this result at better than $8\sigma$, a remarkable conclusion if the introduction 
of $k_{\rm min}$ in the power spectrum turns out to be correct.

For obvious reasons, this outcome is highly relevant to the interpretation
of an acoustic scale because it provides a completely independent measurement
of $z_{\rm cmb}$. At large angles, corresponding to multipoles $\ell\lesssim 30$,
the dominant physical process producing the anisotropies is the Sachs-Wolfe
effect \citep{Sachs1967}, representing metric perturbations due to
scalar fluctuations in the matter field. This effect translates inhomogeneities
of the metric fluctuation amplitude on the last scattering surface into anisotropies
observed in the temperature today.

From the definition of $u_{\rm min}$, it is trivial to see that the maximum
angular size of the Sachs-Wolfe fluctuations is
\begin{equation}
\theta_{\rm max}={2\pi\over u_{\rm min}}\;.
\end{equation}
In the $R_{\rm h}=ct$ Universe, quantum fluctuations begin to form at
the Planck scale with a maximum wavelength
\begin{equation}
\lambda_{\rm max}=\eta\,2\pi R_{\rm h}(z_{\rm cmb})\;,
\end{equation}
where $\eta$ is a multiplicative factor $\sim O(1)$ \citep{MeliaGenova2018}.
Therefore,
\begin{equation}
\ln(1+z_{\rm cmb})=\eta u_{\rm min}\;.
\end{equation}
For example, if $\eta\sim 2/3$, then $z_{\rm cmb} = 17.05^{+8}_{-5}$.
This is a rather significant result because it provides a firm
confirmation that our estimate of $z_{\rm cmb}$ based on the observed
BAO in $R_{\rm h}=ct$ may be correct in the context of this model.
Incidentally, aside from the evidence provided against basic, slow-roll inflation
by the non-zero value of $k_{\rm min}$, the emergence of $\theta_{\rm max}$,
and its implied value of $z_{\rm cmb}$, also introduces significant tension
with the inferred location of the last scattering surface in $\Lambda$CDM based
on the first acoustic peak of the CMB power spectrum. But an extended discussion
concerning this new result is beyond the scope of the present paper,
whose principal goal is an examination of the possible origin of the CMB
in the $R_{\rm h}=ct$ model.

Returning now to Equation~(22), we see that identifying the BAO scale as
the acoustic horizon gives
\begin{equation}
{c_{\rm s}(t_*)\over c/\sqrt{3}}\approx {\beta\over 20}\;.
\end{equation}
As we have seen, part of the reduction of $c_{\rm s}$ below its relativistic
value in $R_{\rm h}=ct$ is due to the fact that $\rho_{\rm r}$ is only $0.2\rho$
in the early Universe. But that still leaves about a factor $4$ unaccounted for
in Equation~(18). Perhaps this is indirect evidence that radiation and dark
energy are coupled strongly during the acoustically active period and that
the sound speed of dark energy cannot be ignored. But without new physics beyond
the standard model, from which such properties would be derived, there is
little more one can say without additional speculation.

\section{Dust vs Recombination in $R_{\rm h}=ct$}
The physical attributes of the LSS that we have just described in the $R_{\rm h}=ct$
universe echo some of the theoretical ideas explored decades ago, though these
were abandoned in favour of a recombination at $z_{\rm cmb}\sim 1080$ scenario.
Before attempting to rescue the dust origin for the CMB, it is essential to
scrutinize globally whether such a proposal makes sense in terms of what we know
today. In general terms, there are at least three observational signatures that
may be used to distinguish between recombination and dust opacity as the origin
of the CMB, and we consider each in turn. In addition, there are
several other potential shortcomings that simply would not work in $\Lambda$CDM,
providing a strong argument {\sl against} the dust model in standard cosmology,
though these are removed quite easily in the context of $R_{\rm h}=ct$, so
that a dust origin for the CMB is virtually unavoidable in this alternative
cosmology. We shall summarize these issues and how they are resolved
in $R_{\rm h}=ct$ at the end of this section.

\subsection{Recombination lines}
The first of these signatures is quite obvious and rests on the expectation
that recombination lines ought to be present at some level in the CMB's spectrum
if the current picture is correct, whereas all such lines would have been
completely wiped out by dust rethermalization. The expectation of seeing
recombination lines from $z_{\rm cmb}$ is so clear cut that extensive simulations
have already been carried out for this process in the context of $\Lambda$CDM
\citep{Rubino-Martin2006,Rubino-Martin2008}. The effect of recombination
line emission on the angular power spectrum of the CMB is expected to be quite
small, of order $\sim 0.1\mu{\rm K}$--$0.3\mu{\rm K}$, but may be separated from
other effects due to their peculiar frequency and angular dependence. Narrow-band
spectral observations with improved sensitivities of future experiments may therefore
measure such deviations if the CMB was produced by recombination.

\subsection{The CMB Spectrum}
A second signature has to do with the CMB's radiation spectrum itself. Clearly,
the opacity in a plasma comprised primarily of Hydrogen and Helium ions and their
electrons is dominated by Thomson scattering, which does not alter the spectral
shape produced at large optical depths as the CMB photons diffuse through the
photosphere. There is, however, the issue of how much dilution of the blackbody
distribution occurs in a scattering medium, which does not alter the `colour'
temperature of the radiation, but reduces its intensity below that of a true
Planck function.

We will not be addressing this specific question here because our primary focus
is dust opacity, which has an alternative set of issues, including the fact that
the efficiency of dust absorption is frequency dependent \citep{Wright1982}.
To address this point, and its impact on the shape of the CMB's radiation spectrum,
let us begin by assuming a density $n_{\rm d}(\Omega,t)$ of thermalizers with a
temperature $T_{\rm d}(\Omega,t)$ at time $t$ and in the direction $\Omega\equiv
(\theta,\phi)$. The efficiency of absorption $Q_{\rm abs}$ (in units of comoving
distance per unit time) of the thermalizers depends on several factors, including
geometry, frequency, composition and orientation.

Then, assuming Kirchoff's law with isotropic emission by each radiating surface along
the line-of-sight, and recalling that the invariant intensity scales as $\nu^{-3}$,
we may write the intensity observed at frequency $\nu_0$ in the direction $\Omega$ as
\begin{eqnarray}
I(\nu_0,\Omega)&=&\langle\sigma\rangle {2h\nu_0^3\over c^2}\int_0^{t_0}\,
dV(t)\;n_{\rm d}(\Omega,t)\times\nonumber\\
&\null&\hskip-0.6in{\langle Q_{\rm abs}(\nu[\nu_0,t])\rangle\over d_L(t)^2}
P(\nu[\nu_0,t],T_{\rm d}[\Omega,t])\,e^{-\tau(\nu_0,\Omega,t)}\,,
\end{eqnarray}
where $\langle\sigma\rangle$ is the average cross section of the thermalizers,
$\langle Q_{\rm abs}\rangle$ is an average over the randomly oriented thermalizers
in the field of unpolarized radiation, $d_L$ is the luminosity distance, $dV$
is the comoving volume element, and
\begin{equation}
P(\nu,T)\equiv {1\over \exp(h\nu/kT)-1}
\end{equation}
is the Planck partition function, so that
\begin{equation}
B(\nu,T)\equiv {2h\nu^3\over c^2}P(\nu,T)
\end{equation}
is the blackbody intensity. In addition, the quantity
\begin{equation}
\tau(\nu_0,\Omega,t)=\langle\sigma\rangle \int_t^{t_0} dt\;
\langle Q_{\rm abs}(\nu[\nu_0,t])\rangle\,n_{\rm d}(\Omega,t)
\end{equation}
is the optical depth due to the thermalizers along the line-of-sight
between time $t$ and $t_0$.

Let us further assume a scaling law
\begin{equation}
n_{\rm d}(\Omega,t)=n_{\rm d}(\Omega,0)(1+z)^\epsilon\;.
\end{equation}
Expressing these integrals in terms of redshift $z$, we therefore have
\begin{eqnarray}
I(\nu_0,\Omega)&=&\tau_0(\Omega){2h\nu_0^3\over c^2}\int_0^\infty\,dz^\prime\,{(1+z^\prime)^{\epsilon-1}
\over c E(z^\prime)}\times\nonumber\\
&\null&\hskip-0.92in \langle Q_{\rm abs}(\nu_0[1+z^\prime])\rangle
P(\nu_0[1+z^\prime]\,T_{\rm d}[\Omega,z^\prime])\,e^{-\tau(\nu_0,\Omega,z^\prime)}\,,
\end{eqnarray}
and
\begin{equation}
\tau(\nu_0,\Omega,z)=\tau_0(\Omega)\, \int_0^z dz^\prime\;
{(1+z^\prime)^{\epsilon-1}\over c E(z^\prime)}\,\langle Q_{\rm abs}(\nu_0[1+z^\prime])\rangle
\end{equation}
where
\begin{equation}
\tau_0(\Omega)\equiv {c\over H_0}\,\langle\sigma\rangle\,n_{\rm d}(\Omega,0)\;,
\end{equation}
and
\begin{equation}
E(z)\equiv {H(z)\over H_0}\;.
\end{equation}

Noting that
\begin{eqnarray}
{d\over dz}e^{-\tau(\nu_0,\Omega,z)}&=&-\tau_0(\Omega){(1+z)^{\epsilon-1}\over
c E(z)}\langle Q_{\rm abs}(\nu_0[1+z])\rangle\nonumber\\
&\null&\times e^{-\tau(\nu_0,\Omega,z)}\;,
\end{eqnarray}
we can see from Equation~(41) that
\begin{eqnarray}
I(\nu_0,\Omega)&=&-{2h\nu_0^3\over c^2}\int_0^\infty\,dz^\prime\,
P(\nu_0[1+z^\prime],T_{\rm d}[\Omega,z^\prime])\nonumber\\
&\null& \times {d\over dz^\prime}e^{-\tau(\nu_0,\Omega,z^\prime)}\;,
\end{eqnarray}
and therefore integrating by parts, we find that
\begin{eqnarray}
I(\nu_0,\Omega)&=&B(\nu_0,T_{\rm d}[0])+{2h\nu_0^3\over c^2}\int_0^\infty\,dz^\prime\,
e^{-\tau(\nu_0,\Omega,z^\prime)}\nonumber\\
&\null&\times{d\over dz^\prime} P(\nu_0[1+z^\prime],T_{\rm d}[\Omega,z^\prime])\;.
\end{eqnarray}

We see that the intensity of the CMB measured at Earth may deviate from that
of a true blackbody, but only if the second term on the right-hand side of
this equation is significant. Notice, however, that regardless of how the
optical depth $\tau(\nu_0,\Omega,z)$ varies with $\nu_0$, there is a strictly
zero deviation from a true Planckian shape for $T_{\rm d}(z)\propto (1+z)$,
which one may readily recognize from Equation~(37). If the dust and the radiation
it rethermalizes near the photosphere (at the LSS) are in equilibrium (see
discussion below concerning what is required to sustain this equilibrium), $T_{\rm d}$
is expected to follow the evolution of the photon temperature (Equation~13) and,
coupled with the fact that $\nu\propto (1+z)$ in all cases, we see that $P(\nu,T)$
is then independent of redshift. Therefore, $(d/dz^\prime)P=0$ in Equation~(47),
leaving $I(\nu_0,\Omega)=B(\nu_0,T_{\rm d}[0])$ at all frequencies \citep{Rowan1979}.

The key issue is therefore not whether the dust opacity is frequency dependent
but, rather, whether the dust reaches local thermal equilibrium with the radiation.
The answer to this question is yes, as long as enough dust particles are generated
to produce optical depths $\tau(\nu_0,\Omega,z)\gg 1$ at $z\sim z_{\rm cmb}$.
Though framed in the context of $\Lambda$CDM, the early work on this topic
already established the fact that a medium could be rendered optically thick
just with dust, even if the latter constituted a mere percentage level density
compared to those of other constituents in the cosmic fluid
\citep{Rees1978,Rowan1979,Wright1982,Rana1981,Hawkins1988}.

In the context of $R_{\rm h}=ct$, we may estimate whether or not this
holds true as follows. Extremely metal-poor stars
have been detected, e.g., in the Galactic bulge \citep{Howes2015}, possibly
revealing a remnant trace of the Pop III stars formed prior to $z\sim 15$.
These data support the conventional picture of an extremely low metal abundance
in the ISM prior to Pop III stellar nucleosynthesis. We do not yet have a
tight constraint on the metallicity between Pop III and Pop II star formation,
but let us parametrize its value relative to solar abundance as $f_{\rm Z}$. We shall
argue in the next subsection that the dust was created prior to $z\sim 16$
and then destroyed by Pop II supernovae at the start of the epoch of
reionization (i.e., $z\sim 15$).

Assuming a Hubble constant $H_0=67.7$ km s$^{-1}$ Mpc$^{-1}$ and a
baryon fraction $\Omega_{\rm b}\sim 0.04$ \citep{Planck2016a},
it is straightforward to estimate the comoving mass density of metals,
$\rho_{\rm s}(z=16)\sim 4\times 10^{-29}f_{\rm Z}$ g cm$^{-3}$ at
$z=16$. Therefore, for a bulk density of $\sim 2$ g cm$^{-3}$ of
silicate grains, and a grain radius $r_{\rm s}\sim 0.1$ micron, the dust
number density would have been $n_{\rm s}(z=16)\sim 5\times 10^{-15}f_{\rm Z}$
cm$^{-3}$. At $z=16$, the CMB spectrum ranged from $\lambda_{\rm min}\sim 0.003$ cm
to $\lambda_{\rm max}\sim 0.02$ cm, for which the dust absorption efficiency
was $Q(\lambda_{\rm min})\sim 0.02$ and $Q(\lambda_{\rm max})\sim 0.003$
\citep{Draine2011}. And therefore the photon mean free path $\langle l_\gamma\rangle$
due to dust absorption is estimated to lie between the limits $\sim 3\times 10^{25}
f_{\rm Z}^{-1}$ cm and $\sim 2\times 10^{26} f_{\rm Z}^{-1}$ cm. By comparison, the
gravitational (or Hubble) radius at that redshift was $R_{\rm h}(z=16)\sim
10^{27}$ cm. Thus, every photon in the CMB would have been absorbed by
dust prior to $z\sim 16$ as long as $f_{\rm Z}\gtrsim 0.2$, i.e., about
$20\%$ of the solar value, which is not at all unreasonable.

Correspondingly, the dust temperature must remain in equilibrium with
the CMB radiation field (see Eq.~43). There are two important factors guiding
this process. The first is based on the average heating $H(T)$ and cooling
$K(T_{\rm d})$ rates for a given dust particle, while the second is due to
the fact that each absorption of a photon produces a quantum change in the
dust particle's temperature that may be strongly dependent on its size
\citep{Weingartner2001,Draine2001}. In the cosmological context,
the dust is heated by an isotropic radiation field with an angle-averaged
intensity $J_\lambda=B(\lambda,T)$ (see Eq.~38), where $T(z=16)\approx 46$ K,
unlike our local neighborhood, where the primary heating agent is UV light.
Thus, a typical dust particle is heated at a rate $H(T)=4\pi r_{\rm s}^2
\int_0^\infty d\lambda\,\pi B(\lambda,T)Q(\lambda)$, in terms of the previously
defined absorption efficiency $Q(\lambda)$. According to Kirchoff's law, its
emissivity is proportional to $B(\lambda,T_{\rm d})Q(\lambda)$, and so its
cooling rate may be similarly written $K(T_{\rm d})=4\pi r_{\rm s}^2
\int_0^\infty d\lambda\,\pi B(\lambda,T_{\rm d})Q(\lambda)$. These integrals
are identical, except when $T_{\rm d}\not=T$.

To gauge how long it would take for the dust to reach equilibrium with the
CMB radiation field if these temperatures were not equal, consider the temperature
evolution equation $C(T_{\rm d}) \;dT_{\rm d}/dt=H(T)-K(T_{\rm d})$, where $C(T_{\rm d})$
is the heat capacity. At $T_{\rm d}\sim 46$ K, $C\sim 0.2\, k_{\rm B}N_{\rm s}$
\citep{Draine2001}, where $k_{\rm B}$ is Boltzmann's constant and $N_{\rm s}$
is the number of molecules in the dust grain. For a $\sim 0.1\mu$m sized particle,
$N_{\rm s}\sim 3\times 10^8$ \citep{Weingartner2001}, so putting
$\langle Q(\lambda)\rangle\sim 0.012$, one finds that $dT_{\rm d}/dt\sim
10^{-7}(T^4-T_{\rm d}^4)$. Thus, assuming that either $H(T)$ or $K(T_{\rm d})$
is dominant, we infer that it would take about $50$ seconds for the dust to
reach equilibrium at $T= T_{\rm d}\sim 46$ K. It is therefore reasonable to
assume that dust was thermalized with the radiation at $z\sim 16$.

The second issue is more constraining. Upon absorbing a photon with
wavelength $\lambda$, a dust grain containing $N_{\rm s}$ molecules undergoes
a change in temperature $\Delta T_{\rm d}=hc/\lambda\,C(T_{\rm d})\sim
7.2\;(\lambda\,N_{\rm s})^{-1}$ K. For the larger grains (i.e., $r_{\rm s}\sim
0.1-0.3$ $\mu$m), with $N_{\rm s}\sim 3\times 10^8-10^{10}$, this is a
minuscule fraction ($\sim 10^{-9}-10^{-8}$) of the equilibrium temperature
$T_{\rm d}=46$ K throughout the wavelength range $\lambda\sim 0.003-0.02$ cm,
so the smooth evolution in $T_{\rm d}$ described in previous paragraphs seems
perfectly attuned to the physics at $z\sim 16$. Smaller grains have less heat
capacity and a reduced radiating area, however, so the absorption of photons
can lead to temperature spikes \citep{Draine2001}. At $r_{\rm s}\sim 0.003$
$\mu$m, we have $N_{\rm s}\sim 1.4\times 10^4$, so $\Delta T_{\rm d}/T_{\rm d}
\sim 6\times 10^{-4}-4\times 10^{-3}$. Evidently, the assumption of a smooth
evolution in $T_{\rm d}$ starts to break down for grains smaller than this,
since they proceed through stochastic heating via absorption and cooling between
the spikes. The dust model required for consistency with the observed spectrum
of the CMB therefore consists of silicates with sizes $\sim 0.003-0.3$ $\mu$m,
or even larger, though for sizes $\gtrsim 0.3$ $\mu$m, we would then violate
our previous estimate of $n_{\rm s}(z=16)$ and the satisfactory result that
$f_{\rm Z}\sim 0.2$.

As modeled here, the dust is optically thick at all relevant
frequencies. But once the dust is destroyed, however, the principal contributor
to the optical depth affecting the CMB spectrum is Thomson scattering within
the ionized medium across the epoch of reionization. At least for this
process, one would not expect a discernible difference between the dust
and recombination models because the structure of the reionization region
is essentially the same in both cases. The observations constrain when
reionization began and ended, and the physics responsible for this process
is essentially independent of the background cosmology.  Certainly, there
are percentage differences arising from the respective age-redshift
relationships, which affect the variation in baryonic density with time,
but a detailed calculation (Melia \& Fatuzzo 2016) has already shown that
the optical depth through this region would be consistent with the value
(i.e., $\tau\sim 0.066$) measured by {\it Planck} \citep{Planck2018}
in both cases.

Finally, let us quantitatively confirm our earlier statement concerning the
negligible impact of Pop III stars on the overall background radiation field.
Much more massive ($500\;M_\odot\gtrsim M\gtrsim 21\;M_\odot$) than
stars formed today \citep{Bromm2004,Glover2004}, Pop III
stars emitted copious high-energy radiation that ionized the halos within
which they formed \citep{Johnson2007}. Following their brief
($\sim 10^6-10^7$ yr) lives, a large fraction of these stars \citep{Heger2003}
exploded as SNe, ejecting the first heavy elements into the
interstellar medium \citep{Whalen2008}. Given the dust size and
required number (see above), we estimate that roughly $9\times 10^{44}$
g Mpc$^{-3}$ (co-moving volume) of dust material needed to be injected
into the interstellar medium during the principal epoch ($20\gtrsim z\gtrsim
15$) of Pop III star formation.

The ultimate fate of the Pop III stars depended on their mass prior to the
SN explosion. For a mass $M\lesssim 40\;M_\odot$, roughly $20\%$
of the mass was ejected into the interstellar medium as metals, leaving
a compact remnant behind. For $M\gtrsim 140\;M_\odot$, the explosion
was much more powerful, dispersing as much as $\sim 50\%$ of the mass
\citep{Heger2002}. For the sake of illustration, let us adopt a
typical mass $M\sim 100\;M_\odot$, with a typical ejection fraction of
$30\%$ (between these two limits). In the $R_{\rm h}=ct$ universe,
 $1+z=1/tH_0$, from which we estimate an interval of time $\Delta t
\sim 200$ Myr between $z=15$ and $20$. Thus, $\sim 1.5\times 10^8$
Mpc$^{-3}$ Pop III stars must have exploded as SNe to provide the
required dust.

Prior to exploding, however, these Pop III stars also injected a copious
amount of radiation into the ambient medium. A typical Pop III star
with mass $M\sim 100\;M_\odot$ was a blackbody emitter with radius
$R_*=3.9\,R_\odot$ and surface effective temperature $T_*=10^5$ K,
so its bolometric luminosity would have been $\sim 4\times 10^{39}$ erg
s$^{-1}$. Thus, the total energy density radiated by these stars
during their lives would have been $U_{III}\sim 4\times 10^{63}$
erg Mpc$^{-3}$. By comparison, the CMB energy density at $z\sim 16$
was $U_{\rm cmb}\sim 8\times 10^{65}$ erg Mpc$^{-3}$.  Evidently,
$U_{III}/U_{\rm cmb}\sim 0.5\%$, a negligible fraction. In terms of
the photon number, this ratio would have been even smaller, given
that the average energy of a photon radiated by the stars was much
higher than that of the CMB.

A somewhat related issue is the nature of the cosmic
infrared background (CIB), and whether it may be related in some way
to a dusty origin for the CMB. Most of the CIB is believed to have been
produced by extragalactic dust at $z\sim 2$ \citep{Planck2011}.
The mechanism for producing the CMB and CIB in this model are, however,
quite different. The CMB in this picture was produced by saturated dust
absorption and emission at $16\gtrsim z\gtrsim 14$, with all of the CMB
photons having been absorbed prior to $z\sim 14$. The dust producing
the CIB at $z\sim 2$ was presumably heated by stars and quasars
near that redshift, thereby producing an infrared signal with a different
temperature profile. The CIB and CMB would have been created under
very different physical conditions, with the high-$z$ component in
thermal equilibrium with the dust, and the lower-$z$ component produced
by dust heated by higher frequency radiation.  As we showed earlier,
dust heating by Pop II and III stars at $16\gtrsim z\gtrsim 14$ was
insignificant compared to the CMB. The reverse situation appears to
have materialized at $z\sim 2$.

\subsection{Frequency-dependent Power Spectrum}
The third crucial signature that may distinguish between dust and recombination
has to do with anisotro\-pies in the temperature distribution across the sky and
how they vary among surveys conducted at different frequencies. In simple
terms, one does not expect photospheric depth effects to determine the observed
distribution of fluctuations in the case of Thomson scattering because the optical
depth is independent of frequency. Thus, maps made at different frequencies should
reveal exactly the same pattern of anisotropies since all of the relic photons
are freed from essentially the same LSS. An important caveat, however,
is that this simplified recombination picture in the standard model may be
ignoring an effect, due to Rayleigh scattering by neutral hydrogen, that itself
could produce a percentage-level dependence of the power spectrum on frequency, as
we shall discuss later in this section.

Assuming that the power spectrum is frequency-independent would
almost certainly not be valid in the case of dust if its opacity also depends on
frequency. Although photospheric depth effects might not significantly change the
shape and size of the larger fluctuations from one map to another, they might alter
the observed pattern of anisotropies on the smaller scales if the angular diameter
distance between the LSS's at two different frequencies is comparable to the
proper size of the fluctuations themselves. These differences would, at some level,
produce variations in the CMB power spectrum compiled at different frequencies.

A detailed analysis of the dependence of the CMB power spectrum on frequency
was reported recently by the Planck collaboration \citep{Planck2016a}, following an initial
assessment of such effects based on the WMAP first-year release in \cite{Hinshaw2003}
(see, e.g., their fig.~2). {\it Planck} maps at different
frequencies constrain the underlying CMB differently and cross-correlating them
is quite challenging, in part due to the changing foreground conditions with
frequency. The {\it Planck} analysis has shown that residuals in the half-mission
TT power spectra clearly do vary from one cross power spectrum to the next,
sampling a frequency range $70-217$ GHz, though this could be due to several
effects, including foreground systematics, as well as possible intrinsic
variations in the location of the LSS. One may also gauge the dependence of
the multipole power coefficients on frequency by varying the maximum multipole
number $\ell_{\rm max}$ included in the analysis, from $\sim 900$ to several
thousand, thereby probing a possible greater variation in the observed
anisotropies on small scales compared to the larger ones. This particular
test produces shifts in the mean values of the optimized cosmological
parameters by up to $\sim 1\sigma$, in ways that cannot always be related
easily to non-cosmological factors. In addition, the cross power spectrum
at lower frequencies ($\lesssim 100$ GHz) shows variations in the amplitude
$D_{\ell}$ of up to $\sim 4\sigma$ compared to measurements at higher
frequencies.

Overall, {\it Planck} finds a multipole power varying an
amount $\Delta D_\ell$ (increasing with multipole number $\ell$ over the
frequency range $\sim 70-200$ GHz) anywhere from $\sim 40\;\mu$K$^2$
at $\ell\sim 400$, to $\sim 100\;\mu$K$^2$ at $\ell\gtrsim 800$. Thus,
with $D_\ell\sim 2000\;\mu$K$^2$ over this range, one infers a maximum
possible variation of the power spectrum---as a result of frequency-induced
changes in the location of the LSS---to be $\sim 2\%$ at $\ell\sim 400$,
increasing to $\sim 5\%$ for $\ell\gtrsim 800$.

Thus, in order for a dust origin of the CMB to be consistent with current
limits, the angular-diameter distance to the LSS cannot vary with frequency so much that
it causes unacceptably large variations in the inferred angular size of the
acoustic horizon. Earlier, we estimated that $z_{\rm cmb}\sim 16$ in the
$R_{\rm h}=ct$ universe. This redshift is interesting for several reasons,
one of them being that it coincides almost exactly with the beginning of
the epoch of reionization at $z\sim 15$. It is tempting to view this as more than a mere
coincidence, in the sense that the ramp up in physical activity producing
a rapid increase of the UV emissivity around that time would not only
have reionized the Hydrogen and Helium, but also destroyed the dust. So
a viable scenario in this picture would have the medium becoming optically thick
with dust by $z\sim 16$, then rapidly thinning out due to the destruction of
the dust grains by $z\sim 15$. Any variation in the location of the LSS
would then be limited to the range of angular-diameter distances between
$z\sim 14-15$ and $16$.

We can easily estimate the impact this would have on the inferred angular
size $\theta_{\rm s}$. Assuming the medium was optically thick at $z_{\rm cmb}$
and that it became mostly transparent by $z=z_{\rm cmb}-\Delta z$, one can
easily show from Equation~(29) that the change in $\theta_{\rm s}$ would be
\begin{equation}
\Delta\theta_{\rm s}={r_{\rm BAO}^{R_{\rm h}=ct}\over R_{\rm h}(t_0)}
\left[{1\over\ln(1+z_{\rm cmb}-\Delta z)}-{1\over\ln(1+z_{\rm cmb})}\right]\;.
\end{equation}
Table~2 summarizes some critical data extracted from this relation.
Given the relatively weak dependence of $d_A^{R_{\rm h}=ct}(z)$ on $z$
at these redshifts, the apparent angular size of the acoustic horizon
changes very slowly. Consequently, even if it took the Universe $50-100$
Myr to become transparent and initiate the epoch of reionization,
the impact on our inferred
CMB power spectrum appears to be no more than a few percent, consistent with
current observational limits.

\begin{table*}
\vskip 0.2in
\center
\centerline{{\bf Table 2.} Dust photospheric depth at the LSS}
\begin{tabular}{cccc}
\hline\hline
$\Delta z$\qquad\qquad&{$\Delta\theta_{\rm s}$}\qquad&\qquad Percentage\qquad&
\qquad{$\Delta t$}\qquad \\
\qquad&{(deg)}\qquad&\qquad of $\theta_{\rm s}$& \qquad {(Myr)}\qquad \\
\hline
1\qquad\qquad & 0.013 &\qquad 2.2$\%$ & \qquad 53 \\
2\qquad\qquad & 0.025 &\qquad 4.2$\%$ & \qquad 100 \\
\hline\hline
\end{tabular}
\end{table*}

Some support for this idea may be found in our current understanding of how
dust is formed and destroyed in the ISM. Though some differences
distinguish nucleosynthesis and mass ejection in Pop III stars from analogous
processes occurring during subsequent star formation, two factors pertaining to
the life-cycle of dust were no doubt the same: (1) that dust principally formed
within the ejecta of evolved stars; and (2) that it was then destroyed much more
rapidly than it was formed in supernova-generated shock waves. These essential
facts have been known since the earliest observation of shock-induced dust
destruction over half a century ago \citep{Routly1952,Cowie1978,Seab1983,Welty2002},
creating a severe constraint on how much
dust can possibly be present near young, star-forming regions. The early-type
stars among them are the strongest UV emitters; they also happen to be
the ones that evolve most rapidly on a time scale of only $10-20$ Myr and then
end their lives as supernovae. The shocks they produce in the ISM result in
the complete destruction of all grains on a time scale
$\lesssim 100$ Myr \citep{Jones1994,Jones1996}.

When this time scale is compared to the results shown in Table~2, the
idea that the Universe transitioned from being optically thick with
dust at $z\sim 16$ to optically thin by $z\sim 14-15$ becomes quite
significant.  There are several links in this chain, however, and maybe
the correlations we have found are just coincidences. But at face value,
there is an elegant synthesis of basic, well-understood astrophysical
principles that work together to provide a self-consistent picture of
how the cosmic fluid might have become optically thick by $z\sim 16$
due to dust production in Pop III stars, followed by an even
more rapid phase of Pop II star formation and deaths. The earliest of
these would have completely destroyed the dust with their supernova-induced
shocks in a mere $\sim 100$ Myr, liberating the CMB relic photons and
initiating the epoch of reionization by $z\sim 14-15$.

To complete the discussion concerning whether or not an
observed frequency-shift in the power spectrum can distinguish between
the recombination and dust models for the CMB using future high-precision
measurements, however, one must also consider the impact of Rayleigh scattering
by neutral hydrogen, which itself may introduce some frequency dependence
on the observed anisotropic structure.

This effect is due to the classical scattering of long-wavelength photons by
the HI dipole, which has an asymptotic $\nu^4$-dependence on frequency.
Since the transition from fully ionized plasma to neutral hydrogen and helium
is not sudden at recombination, higher frequencies of the observed CMB
anisotropies should be Rayleigh scattered by the fractional density of
HI atoms that builds while recombination proceeds (see, e.g., 
\citealt{Takahara1991,Yu2001,Lewis2013,Alipour2015}). But though this
effect can strengthen considerably with increasing frequency, the blackbody
spectrum also falls rapidly, so there are very few photons where Rayleigh
scattering would be most impactful. The above-referenced studies have shown
that the Rayleigh signal is most likely to be observable over a range
of frequencies $200$ GHz $\lesssim \nu\lesssim800$ GHz, producing a
$\lesssim 1\%$ reduction in anisotropy (for both the temperature and
E-polarization) at $353$ GHz.

Nevertheless, a frequency-dependent dust photospheric depth that we have been
discussing in this section may still be distinguishable from the Rayleigh signal
because it is expected to produce $\lesssim 4\%$ variations in the power spectrum
even at frequencies below $\sim 200$ GHz, where the latter is not observable.
As noted earlier, the percentage-level variations suggested by the latest
{\it Planck} observations are observed in the frequency range $\sim 70$ GHz
$-200$ GHz, where the Rayleigh distortions would be $<<1\%$.

\subsection{E-mode and B-mode Polarization}
The three aspects we have just considered---the detection of
recombination lines, the CMB spectrum, and its possible frequency
dependence in the dust model---will feature prominently
in upcoming comparative tests between the recombination and dust
scenarios. But there are several other factors we must consider,
including what the detection (or non-detection) of E-mode and B-mode
polarization can tell us about the medium in which the CMB is produced.

The linear-polarization pattern can be geometrically decomposed into
two rotational invariants, the E (gradient) mode and B (curl) mode
\citep{Kamionkowski1997,Zaldarriaga1997}. In the standard
model, E-mode polarization is produced by Thomson scattering of partially
anisotropic radiation associated with the same scalar density fluctuations
that produce the temperature hot spots. These are longitudinal compression
modes with density enhancements aligned perpendicular to the direction
of propagation, and therefore result in a polarization pattern with zero
curl. Tensor (or gravitational wave) modes, on the other hand, alter the
frequency of the background anisotropic radiation along diagonals to the
propagation vector as they cross the LSS, and the subsequent Thomson
scattering therefore produces a polarization pattern with a non-zero curl.
The detection of B-mode polarization is therefore an important signature
of tensor fluctuations associated with a quantized scalar (possibly
inflaton) field in the early Universe.

As reported by the \cite{Planck2018}, the foreground polarized
intensity produced by dust in the Milky Way is several orders of magnitude
larger than that seen (or expected) in the CMB. Aspherical dust particles
align with an ambient magnetic field and produce both E-mode and B-mode
polarization. But the relative power in these two components is a complicated
function of the underlying physical conditions, notably the strength of the
magnetic field {\bf B} and its structure (i.e., turbulent versus smooth),
and its energy density relative to the plasma density. Many expected to
see a randomly oriented foreground polarization map with equal powers in
the E-modes and B-modes \citep{Caldwell2017}. Instead, the {\it Planck}
data reveal a surprising E/B anisotropy of a factor $\sim 2$ \citep{Planck2018}. 
Equally important, {\it Planck} also reveals a positive TE correlation 
in the dust emission, to which we shall return shortly.

Once the foreground polarization was subtracted, however, the remaining
signal contained only an E-mode pattern and no B-mode that one could
attribute to the CMB. In further analysis, the CMB peaks were stacked,
revealing a characteristic ringing pattern in temperature associated with
the first acoustic peak (on sub-degree scales), and a high signal-to-noise
pattern in the E-mode stack (see, e.g., their fig.~20). This correlation
between the temperature and E-mode anisotropies observed by {\it Planck}
is therefore consistent with the standard picture (see above), supporting
the view that the CMB must have been created by recombination in the
context of $\Lambda$CDM.

But neither the absence of a B-mode in the fore\-ground-subtracted signal,
nor the TE correlation, can yet rule out a dust origin for the CMB in the
alternative scenario we are considering in this paper. The observations are
not yet precise enough, nor is the theoretical basis for dust polarization
sufficiently well established, for us to say for sure whether B-mode
polarization is/should be present in the foreground-subtracted CMB map.

There are two requirements for dust to emit polarized light:
(1) non-sphericity of the dust grains to allow them to spin about an axis
perpendicular to their semi-major axis, and (2) an organized magnetic field
to maintain alignment of the spin axes. We do not know if the earliest dust
grains produced by Population III stellar ejecta were spherical or not, but
our experience with other dust environments suggests this is quite likely.
Insofar as the magnetic fields are concerned, our current measurements suggest
that---if they exist---intergalactic magnetic fields are probably weaker than
those found within galaxies, where $|{\bf B}_{\rm G}|$ is typically $3-4\,\mu$G
\citep{Grasso2001}, but are certainly not ruled out. Observations of
Abel clusters imply field amplitudes $|{\rm B}_{\rm ICM}|\sim 1-10\mu$G, but
beyond that, no firm measurements have yet been made.

High resolution measurements of the rotation measure in high-redshift
quasars hint at the presence of weak magnetic fields in the early Universe.
For example, radio observations of the quasar 3C191 at $z=1.945$
\citep{Kronberg1994} are consistent with $|{\bf B}_{\rm IGM}|\sim 0.4-4\mu$G.
For the Universe as a whole, some interesting limits may be derived using
the ionization fraction in the cosmic fluid and reasonable assumptions
concerning the magnetic coherence length. If one adopts the largest reversal
scale ($\sim 1$ Mpc) seen in galaxy clusters, one concludes that
$|{\bf B}_{\rm IGM}|\lesssim 10^{-9}$ G (see \citealt{Kronberg1994,Grasso2001}, 
and references cited therein). These fields could be
as small as $\sim 10^{-11}$ G, however, if their coherence length is
much larger. Several other arguments add some support to the view that
the primordial $|{\bf B}_{\rm IGM}|$ could have fallen within this range.
Specifically, the galactic dynamo origin for ${\bf B}_{\rm G}$ is not
widely accepted. The main alternative is to assume that the galactic field
${\bf B}_{\rm G}$ resulted directly from a primordial field compressed
adiabatically when the protogalactic cloud collapsed. This would imply a
primordial field strength $\sim 10^{-10}$ G at $z>5$ at the time when
galaxies were forming, consistent with the observational limits derived
from the rotation measures of high-redshift objects \citep{Grasso2001}.

We simply do not know yet what the magnetic-field strength would have
been during the epoch of Pop II and III star formation and evolution.
It is quite possible, e.g., that the magnetic field could have been even
stronger than $|{\bf B}_{\rm IGM}|$ within the halos where the Pop III
stars ejected most of their dust. Of course, such criteria impact whether
or not the dust grains could have been aligned. Some proposed mechanisms
for this process rely on the strength of {\bf B}, but others---such as
mechanical alignment \citep{Dolginov1976,Lazarian1994,Roberge1995,Hoang2012} 
and radiative alignment \citep{Dolginov1976,Draine1996,Draine1997,Weingartner2003,Lazarian2007} 
are not so sensitive. At this stage, it is safe to assume that our experience 
with dust grain alignment and polarized emission in our local neighborhood 
may be insufficient to fully appreciate the analogous process occurring
during Pop III stellar evolution at $z\sim 16$.

But though we have never seen polarized dust emission from the intergalactic
medium, there are several good reasons to suspect that the dust origin for the
CMB described in this paper could nonetheless account for the polarization
constraints already available today. First, the dust producing the CMB
would presumably have been destroyed prior to $z\sim 14$, so the absence
of polarized dust emission from the IGM at $z<14$ is not an indication that
it lacks a magnetic field (see above).

Second, theoretical work on better understanding the characteristics of
dust emission has begun in earnest, mostly in response to these {\it Planck}
observations. We know for a broad range of physical conditions that
the dust polarization fraction is typically $\sim 6-10\%$ (see, e.g.,
\citealt{Draine2009}), not unlike the $\sim 10\%$ fraction measured
in the CMB \citep{Planck2018}.

Third, we now know that the E-mode and B-mode powers depend on several
detailed properties of the dust profile and the background magnetic field
(see, e.g., \citealt{Caldwell2017,Kritsuk2018,Kim2019}).
In fact, it has been known for several decades that an alignment between
the density structures and the magnetic fields generates more E-mode power
than B-mode \citep{Zaldarriaga2001}. In other words, the E/B asymmetry depends
quite sensitively on the randomness of this alignment, such that a higher
degree of randomness produces less E/B asymmetry. Thus, a highly organized
{\bf B} within the halos where the Pop III star dust was expelled would
have produced a large E/B asymmetry. In their analysis, \cite{Caldwell2017} 
considered this dependence in the context of magnetized fluctuations
decomposed into slow, fast, and Alfv\'en magnetohydrodynamic waves, and
showed that E/B could range anywhere from $\sim 2$ (as observed by {\it Planck}
in the Milky Way), to as much as $\sim 20$, when the medium is characterized
by weak fields and fast magnetosonic waves---the conditions one would have
expected for the dust environment at $z\sim 16$ (see their figure 3 for a
summary of these results). Therefore, the current non-detection of B-mode
polarization in the foreground-subtracted CMB signal cannot yet be used
to rule out the dust scenario described in this paper. Ironically, a
future detection of B-mode polarization could be used to either constrain
inflationary models in the context of $\Lambda$CDM, or the underlying
physical conditions in a magnetized dusty environment at $z\sim 16$,
if the scenario developed in this paper continues to be viable.

Finally, {\it Planck} \citep{Planck2018} has confirmed the
existence of a TE correlation in the foreground dust emission (see above),
suggesting that the overlap seen in the temperature and E-mode stacks
of the foreground-subtracted CMB signal could either have been due to
Thomson scattering in the recombination scenario, or to the polarized
dust emission at $z\sim 16$.

\subsection{Other Potential Shortcomings of the Dust Model}
Lensing of the CMB has been measured with very high precision,
and appears to be consistent with the transfer of radiation over
a comoving distance extending from $z\sim 1080$ to $0$ (for
an early review, see \citealt{Lewis2006}). The latest Planck
data \citep{Planck2016b} would therefore not support
a CMB originating at $z\sim 16$ in the context of $\Lambda$CDM.

But structure formation happened differently in $R_{\rm h}=ct$,
and measures of distance deviate sufficiently from one model
to the next that weak lensing calculations need to be carefully
redone. We do not yet have a complete simulation of the
fluctuation growth in this model over the entire cosmic history,
though some initial steps have been taken \citep{Melia2017a,Yennapureddy2018}.
Insofar as lensing is concerned, there are several key
factors that one may use to qualitatively assess how the lensing effects 
in $R_{\rm h}=ct$ would differ from those in $\Lambda$CDM. Although
the LSS redshift is different in the two models, and the time-redshift
relationship varies by factors of up to $\sim 2$, what matters most
critically in determining the lensing effects are: (1) the comoving
distance to the LSS, (2) the potential well sizes, and (3) the background
pattern of anisotropies at the LSS. 

Together with estimates of the BAO scale (see Table~1 above), the initial 
calculations completed thus far for the formation of structure in $R_{\rm h}=ct$
inform us that the typical potential well size in this model is about $265$ Mpc
(compared with $\sim 300$ Mpc in $\Lambda$CDM), while the comoving distance 
between $z\sim 16$ and $0$ is $\sim 12,200$ Mpc. Thus, one may estimate that
the approximate number of potential wells traversed by the radiation from 
where the CMB originates to $z=0$ is approximately $46$. As it turns out, 
this is almost exactly the same number as in the standard model from 
$z\sim 1080$ to $0$ \citep{Lewis2006}.

The deflection angle due to weak lensing from $z\sim 16$ to $0$ in 
$R_{\rm h}=ct$ is therefore expected to be quite similar
to that from $z\sim 1080$ to $0$ in $\Lambda$CDM. We may estimate
it by assuming that the potentials are uncorrelated, so that the 
total deflection angle should be $\sim 10^{-4}\sqrt{46}$ radians, in 
terms of the approximate deflection angle due to a single well. Thus 
the overall deflection angle is about $2$ arcmins in both models. The 
actual calculation of the remapping of the CMB temperature due to weak 
lensing is much more complicated than this, of course, but the fact that
the scales are so similar suggests that the observed lensing features
probably do not rule out a dust origin for the CMB in $R_{\rm h}=ct$.

Finally, there would be a problem growing the fluctuation amplitude
of $\sim 10^{-5}$ from $z\sim 16$ to $0$ in the standard model,
given that there is barely enough time to do so starting from $z\sim 1080$.
While this is true in $\Lambda$CDM, the time-redshift relationship
in $R_{\rm h}=ct$ is sufficiently different to compensate for the shorter
redshift range. Again, we do not yet have a complete history of the
fluctuation growth in this model, but the growth equation differs from
that in $\Lambda$CDM, primarily because the background metric
is not the same \citep{Melia2017a}. The principal issue, though, is the
timeline $t(z) = t_0/(1+z)$. It is easy to see that the time elapsed
from $z\sim 17$ to today is about $13$ Gyr. By comparison, the
time elapsed since $z\sim 1080$ in $\Lambda$CDM is about
$13.7$ Gyr. One would not claim that this difference is sufficient
to create a problem for $R_{\rm h}=ct$, especially since the two
growth equations are not the same.

\section{Discussion}
Our principal goal in this paper has been to demonstrate how the
zero active mass condition (i.e., $\rho+3p=0$) underlying the
$R_{\rm h}=ct$ cosmology guides the evolution in $\rho_{\rm r}$,
$\rho_{\rm m}$, $\rho_{\rm de}$ and $T(z)$, particularly at early
times when the CMB was produced. Together with additional constraints
from the measured values of $\theta_{\rm s}$ and $r_{\rm BAO}$, and
the adoption of the acoustic horizon as a standard ruler, we have
concluded that $z_{\rm cmb}$ in this model must be much smaller than
its corresponding value in $\Lambda$CDM, eliminating the possibility
that the `recombination' of protons and electrons could have liberated
the CMB relic photons in this model. Finding an alternative mechanism
for producing the CMB in this picture does not have as much
flexibility as one might think, however, because the
physical attributes of the LSS and the measured values of $H_0$
and $T_0$ point back quite robustly to the dust model proposed
several decades ago. Ironically, many of the features in this
model that were resoundingly rejected in the context of $\Lambda$CDM
become fully self-consistent with each other and the data when
viewed with $R_{\rm h}=ct$ as the background cosmology. The fact
that the creation of the CMB at $z_{\rm cmb}\sim 16$ coincides
very well with the onset of the epoch of reionization at $z\sim 15$
is a strong point in its favour, because the astrophysics of
this process is well understood in the local Universe, from which
one expects a correlation between the rapid increase in UV
emissivity and the rapid destruction of dust grains in
star-forming regions.

Fortunately, the observational differences between the recombination
and dust scenarios should be quite distinguishable using the
improved sensitivities of future experiments, thus allowing
us to definitively rule out one or the other of these mechanisms
in the near future. This result may come either (i) from the
detection of recombination lines at $z\sim 1080$, which without
any doubt would rule out dust and very strongly affirm the
recombination model in $\Lambda$CDM, or (ii) affirm a robust
frequency dependence of the CMB power spectrum, with $\sim 5\%$
variations arising from the displacement of the LSS from $z\sim 16$
to $z\sim 15$ (or lower) across the sampled frequency range.

In the meantime, there is much to do on the theoretical front.
Our initial investigation into how acoustic waves might
have evolved in the early $R_{\rm h}=ct$ universe, eventually
producing the multi-peak structure in the temperature spectrum
of the CMB and, later, also the characteristic BAO distance scale
in the distribution of galaxies and the Ly-$\alpha$ forest, has
resulted in a self-consistent picture for the redshift dependence
of the components in the cosmic fluid. By no means should this
study yet be viewed as compelling, however, given that the physics
of fluctuation growth throughout this period is very complex and
dependent on many assumptions, some reasonably justified, others
subject to further scrutiny.

The corresponding picture in the standard model has undergone
several decades of development, based on a combination of simple
arguments---such as the use of Equation~(16) to estimate the
acoustic horizon in the comoving frame---and much more elaborate
semi-analytic and full numerical simulations to follow the various
epochs of halo growth as the dominant contributions to the cosmic
fluid transitioned from radiation to coupled baryon-radiation
components, and finally to matter. In addition, one must introduce
some reasonable distinction between the fluctuations themselves
and the smooth background.

For example, a principal concern with the modeling of growth
across the epoch of recombination is the delayed condensation of
baryons relative to dark matter \citep{Yoshida2003,Naoz2006}.
Structure formation in
the early Universe begins with the gravitational amplification
of small seed fluctuations, which is believed to form dark-matter halos.
Subsequent hydrodynamic processes allow the baryonic gas to fall
into these potential wells, undergoing shock heating and radiative
cooling along the way. Several different studies have indicated
that a substantial difference may therefore exist in the
distribution of baryons and dark matter at decoupling (for
some of the pioneering work on this topic, see
\citealt{Hu1995,Ma1995,Seljak1996,Yamamoto2001,Singh2002}.
After recombination, when the baryons were no longer coupled to
the radiation, gravitational infall caused the baryon density
fluctuations to catch up to the dark matter anisotropies, though
perturbation modes below the Jeans length were presumably delayed
as a result of the initial oscillations.

It is quite clear from this brief overview that the development
of an acoustic scale, and its subsequent evolution throughout the
formation of large-scale structure, not only depends on rather
complex physics, but must also probably vary between
different cosmological models. {\it For example, a principal
difference between the recombination and dust scenarios is
that decoupling in the latter would have occurred well before
the liberation of the CMB relic photons, which means that the
emergence of an acoustic horizon would be mostly hidden from
view by the large dust opacity at smaller redshifts.} The only
features that would have survived across $z_{\rm cmb}$ are
$\theta_{\rm s}$ and the scale $r_{\rm BAO}$, but note that
both of these quantities would have been set well before
$t_{\rm cmb}$, creating some observational ambiguity about
the value of $z_{\rm dec}$, which equals $z_{\rm cmb}$ in
$\Lambda$CDM, but not in $R_{\rm h}=ct$.

This paper represents merely the first step, basically the use
of various measurements to estimate the physical conditions
prior to $z_{\rm cmb}$. And though the picture is self
consistent thus far, some of the essential elements may change,
perhaps considerably, once realistic simulations are carried out.
Nonetheless, the empirical estimate shown in Equation~(35) is quite
robust, because regardless of how and when the baryonic structure
started to form, this average sound speed is required by the
assumed equality of the measured acoustic radius $r_{\rm s}$ and
the BAO scale $r_{\rm BAO}$. This estimate therefore includes
effects, such as oscillations in the coupled baryon-radiation
fluid and the subsequent baryonic catch up. In other words,
we don't actually need to know specifics about the medium
through which the waves propagated to get this number because,
at this level, it is derived from the observations.

As we look forward to further developments in this analysis,
there are several clues and indicators that are already quite evident.
The self-consistent picture emerging in this paper requires a
continued coupling between the various components in the (cosmic)
background fluid, as one may infer directly from figure~1. In the
early universe, the background radiation, dark energy and matter
would necessarily have been coupled. Much of this requires new
physics, but this situation is hardly new or unique. It is difficult
to avoid such a conclusion in any cosmological model. Even
$\Lambda$CDM has several such requirements that are yet to
be resolved. Consider that we have no idea what the inflaton field
is. Yet without it, $\Lambda$CDM cannot resolve the horizon problem.
We also have little idea of how baryonic and dark matter were generated
initially. Certainly, matter was not present at the big bang,
nor during the inflationary phase. Dark energy remains a big
mystery, particularly if it is really a cosmological constant,
given that its density is many orders of magnitude smaller than
quantum field theory requires for the vacuum. All of these issues
await a possible resolution in physics beyond the standard model.

The situation is somewhat different with the fluctuations
themselves. Recent work with this model \citep{Melia2017a} suggests that,
in spite of this coupling, dark energy remained a smooth background,
and did not participate in the fluctuation growth. It is therefore
reasonable to expect that the sequence of dark matter condensation
followed by baryonic catch up (required in $\Lambda$CDM) carries
over in an analogous fashion to $R_{\rm h}=ct$. But there are
several important differences, one of them being that neither
radiation nor matter could apparently have represented more
than $\sim 20-30\%$ of the total energy density at any given
time. This would almost certainly have slowed down the rate of
growth in the early universe, but there would have been ample
time to accommodate this difference given that $t_{\rm cmb}$
in this model is $\sim 849$ Myr, compared to $\sim 380,000$
yr in $\Lambda$CDM.

Aside from the smaller fractional energy density representation
of matter and radiation, there is the additional difference compared
to $\Lambda$CDM brought about by the implied evolution of $\rho_{\rm m}$,
$\rho_{\rm r}$ and $\rho_{\rm de}$ (see figure~1) consistent with
the zero active mass condition. The latter leads to a growth
rate equation for the fluctuations lacking a gravitational growth
term to first order (Melia 2017a). This feature has actually been
quite successful in accounting for the observed growth rate at
$z\lesssim 2$, matching the inferred value $f\sigma_8(0)$ at
redshift zero quite well. By comparison, the corresponding equation
in $\Lambda$CDM predicts a curvature in this rate as a function
of redshift that is not supported by the observations \citep{Melia2017a}.
This growth characteristic in $R_{\rm h}=ct$ also applies
to the early universe, adding to our expectation that the
gravitationally-induced growth of fluctuations was slower
in this model compared to $\Lambda$CDM.

\section{Conclusion}
The acoustic scale associated with the propagation of sound waves prior
to recombination has become one of the most useful measurements in
cosmology, providing a standard ruler for the optimization of several
key parameters in models such as $\Lambda$CDM. The standard model,
however, does not fit the measured BAO scale very well. And a more
recent analysis of the CMB angular correlation function provides
some evidence against basic, slow-roll inflation, making it more
difficult to understand how the horizon problem may be avoided in
$\Lambda$CDM. Given the success of the alternative cosmology known
as $R_{\rm h}=ct$ in accounting for a diverse set of observational
data, we have therefore sought to better understand how the
origin of the CMB could be interpreted in this model.

We have found that the characteristic length ($\sim$$131\pm 4.3$ Mpc)
inferred from large-scale structure may be interpreted as a BAO scale
in $R_{\rm h}=ct$, as long as $z_{\rm cmb}\sim 16$, which would mean
that the location of the LSS would essentially coincide with the
onset of the epoch of reionization. This picture is consistent with the evolutionary
requirements of the zero active mass condition and with our
understanding of the life cycle of dust in star forming regions.

Of course, much work remains to be done. The results look promising
thus far, suggesting that finding a more complete solution,
incorporating the necessary physics to account for the growth of
fluctuations up to $z_{\rm dec}$ and their continued evolution
towards $z_{\rm cmb}$, is fully warranted. This effort is currently
underway and the outcome will be reported elsewhere. On the observational
front, the recombination and dust models for the origin of the CMB
should be readily distinguishable with upcoming, higher sensitivity
instruments, which should either detect recombination lines at
$z\sim 1080$, or establish a robust variation with frequency of the
CMB power spectrum due to the displacement of the LSS from
$z\sim 16$ to $z\sim 14-15$ across the sampled frequency range
at the level of $\sim 2-5\%$.

\acknowledgments
I am grateful to the anonymous referee for several
helpful suggestions to improve the presentation in the manuscript.
I am also very happy to acknowledge helpful discussions with Daniel Eisenstein
and Anthony Challinor regarding the acoustic scale, and with Martin Rees,
Jos\'e Alberto Rubino-Martin, Ned Wright and Craig Hogan for insights
concerning the last-scattering surface. I thank Amherst College for its
support through a John Woodruff Simpson Lectureship, and Purple Mountain
Observatory in Nanjing, China, for its hospitality while part of this work
was being carried out. This work was partially supported by grant 2012T1J0011
from The Chinese Academy of Sciences Visiting Professorships for Senior
International Scientists, and grant GDJ20120491013 from the Chinese State
Administration of Foreign Experts Affairs.

\end{document}